\documentclass[aps,prl,reprint,longbibliography,superscriptaddress,dvipsnames]{revtex4-2}
\usepackage[colorlinks=true,allcolors=blue]{hyperref}
\usepackage{amsmath,amssymb,graphicx,enumitem,braket,bm,bbm,times,xcolor,xfp}

\newcommand{\bs}{\boldsymbol}

\usepackage{tcolorbox}
\newtcolorbox{mybox}{colback=red!5!white,
colframe=red!75!black}

\begin{document}
\title {Cavity magnonics with domain walls in insulating ferromagnetic wires}
\author{Mircea Trif}
\affiliation{International  Research  Centre  MagTop,  Institute  of  Physics,  Polish  Academy  of  Sciences, Aleja  Lotnikow  32/46,  PL-02668  Warsaw,  Poland}
\author{Yaroslav Tserkovnyak}
\affiliation{Department of Physics and Astronomy, University of California, Los Angeles, CA 90095, USA}
\affiliation{Bhaumik Institute for Theoretical Physics, University of California, Los Angeles, CA 90095, USA}
\date{\today}

\begin{abstract}
 Magnetic domain walls (DWs) are topological defects that exhibit robust low-energy modes that can be harnessed for classical and neuromorphic computing. However, the quantum nature of these modes has been elusive thus far. Using the language of cavity optomechanics,  we show how to exploit a geometric Berry-phase interaction between the localized DWs and the extended magnons in short ferromagnetic insulating wires to efficiently cool the DW to its quantum ground state or to prepare nonclassical states exhibiting a negative Wigner function that can be extracted from the power spectrum of the emitted magnons. Moreover, we demonstrate that magnons can mediate long-range entangling interactions between qubits stored in distant DWs, which could facilitate the implementation of a universal set of quantum gates.  Our proposal relies only on the intrinsic degrees of freedom of the ferromagnet, and can be naturally extended to explore the quantum dynamics of DWs in ferrimagnets and antiferromagnets, as well as quantum vortices or skyrmions confined in insulating magnetic nanodisks.

\end{abstract}

\maketitle

{\it Introduction.}|Magnetic domain walls (DWs) are stable topological magnetic configurations that are robust against local deformations \cite{zang2018topology}. Because of their topological stability, capacity for high-density packing, and flexible design, magnetic DWs in ferromagnetic nanowires have emerged as a promising avenue for data storage and information processing in spintronic devices \cite{Beach.2005,Hayashi.2008,Parkin08Science,Grollier2020,KUMAR20221}. Understanding their dynamics and motion is crucial for achieving high-speed operation and efficient manipulation of spintronic devices. 

Miniaturization naturally raises the question of the importance of quantum effects in future DW-based devices. Indeed, several works have proposed recently to harness their quantum properties
in order to encode and process quantum information at the nanoscale \cite{TakeiPRB17,TakeiPRB18,StampsPRB18, PsaroudakiPRL21,PosskePRR22,zou2022domain}. However, such endeavors require the DW to be in the ground state, or at least near its ground state. Until now, no such achievement has
been demonstrated experimentally, largely because the required operation temperatures are too low ($\sim m$K). This raises the question of whether the quantum regime can be reached in practice and, if so, under what conditions? In this work, we answer these questions affirmatively by proposing a ground-state cooling method that exploits the intrinsic magnons (or spin waves) in insulating ferromagnetic devices hosting DWs. 

\begin{figure}[t]
\includegraphics[width=0.95\linewidth]{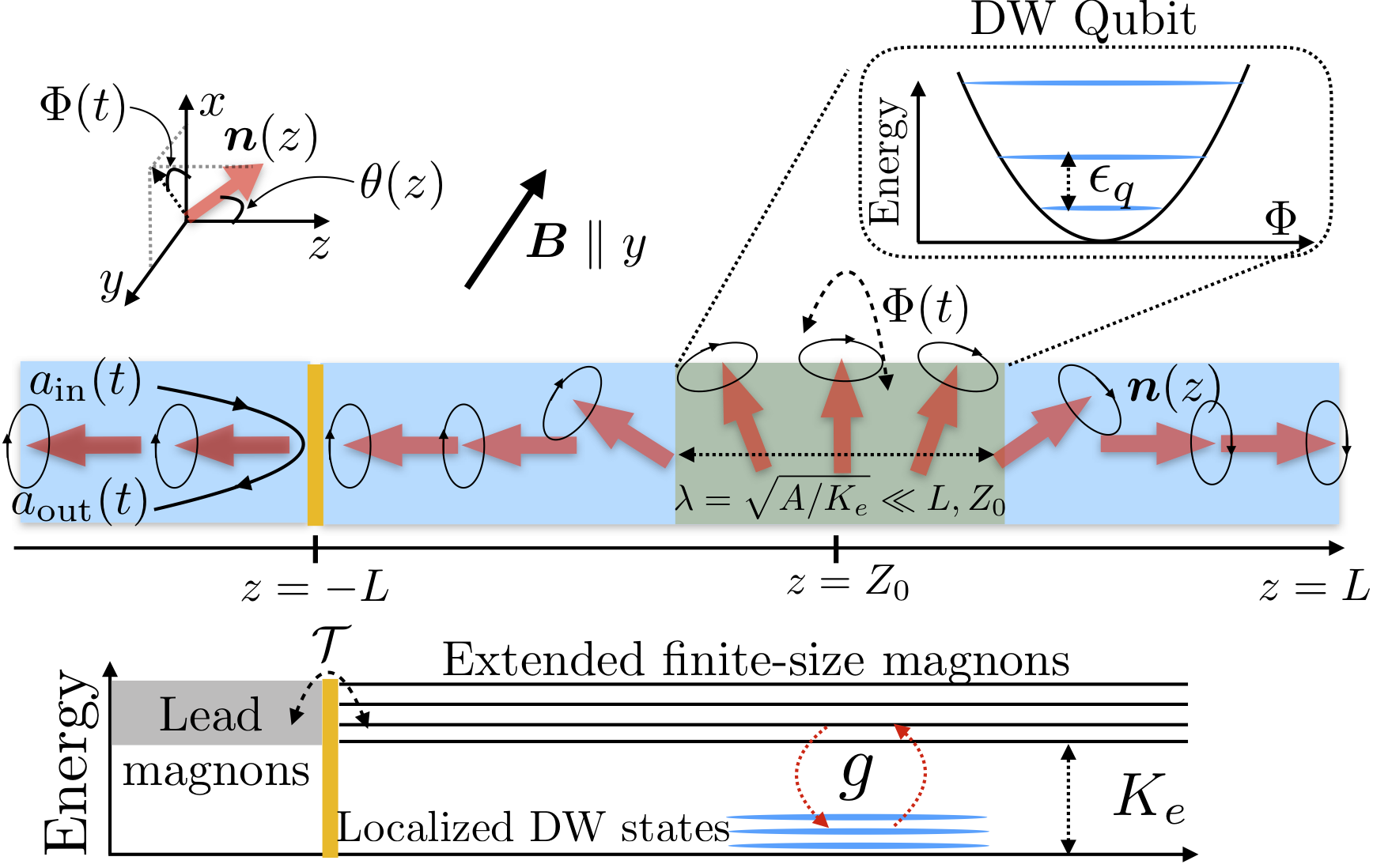}
\caption{Magnonic cavity with a DW: a ferromagnetic insulator wire of length $2L$ and magnetization profile ${\bs n}(z)$ harbors a DW of size $\lambda\ll L$ pinned at position $Z_0$ in the wire. An external magnetic field ${\bs B}\parallel y$ controls the collective angle $\Phi$, allowing to encode a qubit with energy $\epsilon_{q}$ in the lowest two levels. The magnons gapped by the easy $z$-axis anisotropy $K_e$ couple to the DW with a coupling strength $g$ by altering its collective Berry phase. Coupling the cavity magnons to a magnonic waveguide facilitates the manipulation [by the input magnons $a_{\rm in}(t)$] and the detection [from the power spectrum of the output magnons $a_{\rm out}(t)$] of the DW levels.}
\label{fig1} 
\end{figure}

We focus on a finite ferromagnetic insulating wire of length $2L$ and easy-axis anisotropy $K_e$ that harbors one (or several) DW(s) of size $\lambda\ll L$ along its length, as depicted in Fig.~\ref{fig1}. When the DW is pinned in the $z$ direction, the low-energy spectrum is determined by the dynamics of the collective coordinate $\Phi$, which parameterizes the angle of the spin texture of the DW \cite{TataraDW08}. The resulting (quantum) level structure can be manipulated by an external magnetic field ${\bs B}$ that modifies the confining potential $U(\Phi)$ (see Fig.~\ref{fig1}),  rendering it highly anharmonic. Therefore, the two lowest energy levels localized at the DW, $\{|1\rangle,|2\rangle\}$, can be used to encode a qubit below the gapped magnonic levels \cite{TakeiPRB18,zou2022domain}. However, the external magnetic field, in-plane anisotropies, and -- most importantly -- the DW rotational motion induce interactions between the localized DW states and the gapped, extended magnons, which in turn alter their collective dynamics. Instead of viewing this as a nuissance, here we harness such interactions to control the DW dynamics. Using the theoretical framework of cavity optomechanics \cite{MarquardtRMP14,StampsPRB18, BauerPRL18, MarquardtPRB18, StampsPRB21}, we find that the large anharmonicity of the DW (supplemented by dissipation) in the presence of coherently driven magnons enables means of controlling the motion of DW at the quantum level. Specifically,  we advance the conditions for efficient cooling to the ground state as well as preparing stationary DW states that exhibit a negative Wigner function -- a hallmark of a pure quantum regime. Experimentally, such features can be extracted from the power spectrum of the emitted magnons.

{\it Phenomenology of DW-magnon dynamics}.|To grasp the essentials of our proposal, we start with a heuristic discussion of our main findings, relegating the technical details to later sections and the Supplementary Material (SM) \cite{SM}. A DW is described by the collective (and canonically conjugate) coordinates $Z$ and $\Phi$ \cite{TataraDW08}, denoting its position (scaled by $\lambda$) and the rotation angle in the wire, respectively.  In the presence of a strong pinning potential $V_{\rm pin}(Z)=\delta K_e(Z-Z_0)^2/2$, with $\delta K_e$ being its strength, the motion along $z$ is quenched and the low-energy dynamics of a DW with spin $S_{dw}$ (per length $\lambda$) is described by the following rotational Hamiltonian ($\hbar=1$):
\begin{align}
H_{\Phi}&=\frac{1}{2M_{\Phi}}P_{\Phi}^2+U(\Phi)\,.
\label{DWham}
\end{align}
Here, $M_\Phi=(2S_{dw})^2/\delta K_e$ and $P_\Phi$ are the mass of the DW and the canonical momentum associated with rigid rotations of the DW, respectively, while $U(\Phi)$ describes the potential that depends on the magnetic anistrotropies and the external magnetic field. The latter can adjust $U(\Phi)$ from a double-well potential to a single-well potential that is harmonic in nature \cite{zou2022domain}. Promoting the momentum to a quantum variable, i.e.,  $P_{\Phi}\rightarrow-i\partial_\Phi$, allows to solve the Schrodinger equation $H_{\Phi}\psi_{j}(\Phi)=\epsilon_j\psi_{j}(\Phi)$, with $\psi_{j}(\Phi)$ and $\epsilon_j$ being, respectively, the eigenstate and energy of mode $j=1,2,\dots$ of the DW. Importantly, the anharmonicities facilitate the implementation of a qubit in the subspace spanned by the two lowest energy levels with splitting $\epsilon_q\equiv\epsilon_2-\epsilon_1$ \cite{TakeiPRB18,zou2022domain}. 

\begin{figure}[t]
\includegraphics[width=\linewidth]{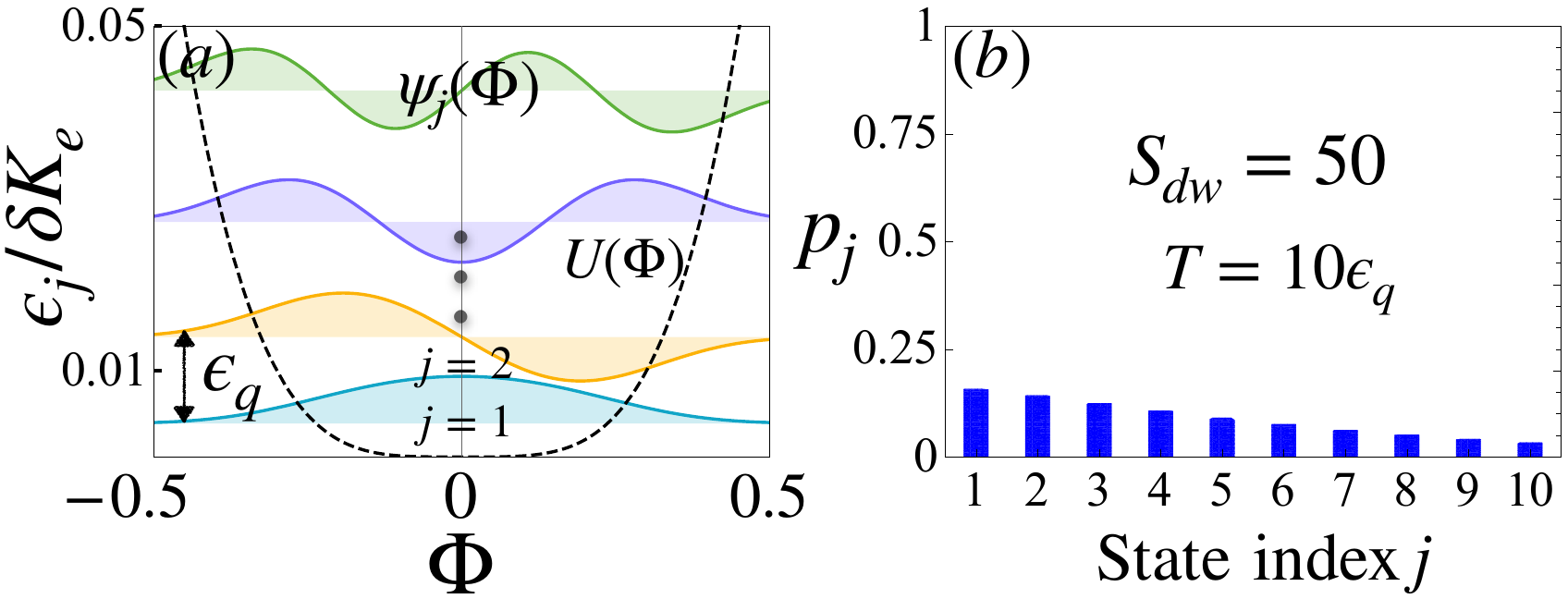}
\caption{(a) The DW energy levels  $\epsilon_j$ and wave-functions $\psi_j(\Phi)$, with $j=1,2,\dots$, for a reduced magnetic field $\tilde{b}=\pi b/2K_h=1$ leading to a quartic effective potential $U(\Phi)$ (dashed line). (b) Occupation of DW levels for $T=10\epsilon_q$, indicating a highly thermal (or classical) state.  The other parameters are $\widetilde{K}_h=5\delta K_e$ and $S_{dw}=50$, while all energies are expressed in terms of $\delta K_e$.}
\label{fig2} 
\end{figure}

When the external magnetic field is adjusted to make the potential quartic, $U(\Phi)\approx\widetilde{K}_h\Phi^4/4$, both the qubit splitting and the anharmonicity are large, $\widetilde{K}_h$ being an effective DW hard-axis anisotropy strength in the $y$ direction \cite{SM}. The resultant spectrum and wave functions are shown in Fig.~\ref{fig2}a. We can gain further insight into the anharmonicity in this limit by using the quasiclassical Bohr-Sommerfeld quantization condition \cite{SM}, to find $\epsilon_j\approx\epsilon_1(j-1/2)^{4/3}$, with $j=1,2,\dots$ and
\begin{align}   
\epsilon_1\approx\frac{0.86} {S_{dw}^{4/3}}\left(\widetilde{K}_h\delta K_e^2\right)^{1/3}\,.
\end{align}
To provide some estimates, for a DW with $S_{dw}\approx 50$, $\widetilde{K}_h=5\delta K_e$, and $\delta K_e\approx 10$ GHz, we find a qubit splitting $\epsilon_q\approx\times10^{-2}\widetilde{K}_h\approx 100$ MHz. We stress that it is possible to achieve larger qubit splittings by selecting ferrimagnetic wires that are close to the compensation point, $S_{dw}\approx0$ \cite{zou2022domain}.  

The decoherence of the DW quantum levels can be accounted for by substituting $H_{\Phi}\rightarrow H_{\Phi}-\Phi\xi(t)$, where $\xi(t)$ is the Langevin force obeying the fluctuation-dissipation condition $\langle\xi(t)\xi(0)\rangle=\alpha S_{dw}\int (d\omega/2\pi)e^{-i\omega t}\omega\coth(\omega/2k_BT)$ at temperature $T$, with $k_B$ being the Boltzmann constant and $\alpha$ the Gilbert damping \cite{TakeiPRB17}.  In a quantum description, it corresponds to a bosonic bath that discriminates between the emission and absorption processes. The transition rate $\gamma_{jk}$ from state $j$ to state $k$ can be evaluated using Fermi's golden rule \cite{BreuerBook}, giving $\gamma_{jk}=\alpha S_{dw}|\langle j|\Phi|k\rangle|^2\epsilon_{jk}[1+\coth(\epsilon_{jk}/2k_BT)]$,
with $\epsilon_{jk}=\epsilon_j-\epsilon_k$, while $\gamma_{kj}=\gamma_{jk}\exp(-\epsilon_{jk}/k_BT)$, in thermal equilibrium.  Using the same qubit parameters, we find $\gamma_{12}\sim\gamma_{21}\approx 10^{-3}\epsilon_q$ for $T=0.05{\rm K} (\approx 10\epsilon_q)$. Although the qubit exhibits low relaxation rates,  the occupation of the DW levels at $T=0.05$ K is highly thermal,  as shown in Fig.~\ref{fig2}b, thus far from its ground state, as desired for qubit initialization.

ext, we qualitatively discuss the interaction with the magnons, fluctuations around the localized DW that extend throughout the wire \cite{KimPRB18}, which can be used to cool the DW towards its ground state.     
In a frame rotating with the DW, the magnons are subject to a fictitious magnetic field pointing along the wire axis $B_{\rm m}(Z_0)\propto \dot{\Phi}{\rm m}_{dw}$, with ${\rm m}_{dw}=(N_\leftarrow-N_\rightarrow)/N_{\rm tot}\equiv Z_0/L$ being the spin up/down fraction of the magnon (which moves ballistically across the DW), $N_{\leftarrow(\rightarrow)}$ the number of spins pointing along the left (right) direction, while $N_{\rm tot}=N_{\leftarrow}+N_{\rightarrow}$ \footnote{This result is strictly valid only for magnons with wavevectors satisfying $kL\gg1$; for magnons with $kL\sim1$ this simple picture remains qualitatively correct, but it is affected slightly by interference effects, as demonstrated in \cite{SM}}. Its physical manifestations are revealed when considering the proximity to a magnetic lead at the wire boundary (see Fig.~\ref{fig1}), with a coupling that picks a term $\exp(i\Phi)$ in a frame rotating with the DW. This latter contribution can be removed by a gauge transformation, resulting in a net magnetic field $B_{\rm m}\rightarrow B_{\rm m}'\propto\dot{\Phi}(1-{\rm m}_{dw})$ acting on the magnons. In a Hamiltonian description, the velocity coupling corresponds to a vector potential that acts on the magnons \cite{SM}, with a maximum (minimum) strength at $Z_0=-L$ ($Z_0=L$). If a specific magnonic mode of energy $\omega_n$ is triggered by an external driving of frequency $\Omega_n$ and strength $h_n$ at the left end, only this mode effectively participates in the dynamics. Then, we arrive at the following DW-magnon coupling Hamiltonian:
\begin{align}
    \!\!\!\!H_{\rm tot}&\approx \frac{(P_\Phi-P_na_n^\dagger a_n)^2}{2M_\Phi}+\Delta_na_n^\dagger a_n+h_n(a_n^\dagger+a_n)\,,
\label{DW_magnons}    
\end{align}
where $a_n$ ($a_n^\dagger$) are the annihilation (creation) operator for the magnon mode with energy $\Delta_n\equiv\omega_n-\Omega_n$, while  $P_{n}\equiv P_{n}(Z_0)=1-{\rm m}_{dw}$ quantifies the Berry-phase mediated DW-magnon coupling \cite{remark}. Eq.~\eqref{DW_magnons} is formally analogous to an optomechanical system, with the DW angle and the magnons replacing the mechanical position and the cavity photons, respectively \cite{MarquardtRMP14}. Note that in the absence of the drive ($h_n=0$), the coupling is a pure gauge shift and therefore inconsequential. Eq.~\eqref{DW_magnons} represents one of our main findings, and its heuristic derivation is complemented by a microscopic calculation in the next section, and further technical details are provided in the SM \cite{SM}. Importantly, Eq.~\eqref{DW_magnons} is general and universal (since it is rooted in Berry-phase effects), and it is the starting point for unraveling the quantum dynamics of the macroscopic DW states.

{\it Microscopic model.}|We assume that the ferromagnetic wire is well below the Curie temperature $T_c$, such that the long-wavelength dynamics can be described in terms of coarse-grained continuous fields ${\bs n}(z)=\{\sin{\theta}\cos{\phi},\sin{\theta}\sin{\phi},\cos{\theta}\}$. The Lagrangian for the magnetic wire reads \cite{TataraDW08}:
\begin{align}
\mathcal{L}_w[{\bs n},\dot{\bs n}]&=-\int_{-L}^L dz\left(s\frac{{\bs d}\times{\bs n}}{1+{\bs n}\cdot{\bs d}}\dot{\bs n}+\mathcal{H}_F[{\bs n}]\right)\,,\nonumber\\
\mathcal{H}_F[{\bs n}]&=\frac{1}{2}[A(\partial_z{\bs n})^2-K_en_z^2+K_hn_y^2+b n_y]\,,
\label{lagrange}
\end{align}
where ${\bs d}$ is the  (arbitrary) direction of the Dirac string, while $s$, $A$, $K_e$,  $K_h$, and $b$ are the spin density, exchange (stiffness), the easy-axis anisotropy, the hard-axis anisotropy, and the magnetic field, respectively. To account for dissipation consistent with the Landau-Lifhitz-Gilbert (LLG) equation obeyed by the magnetization dynamics, the above Lagrangian can be supplemented with the Rayleigh function $\mathcal{R}[\dot{\bs n}]=\alpha s\int dz\dot{\bs n}^2(z)/2$. Finally, the coupling between the wire and the magnetic lead reads $H_{\rm c}=\mathcal{T}{\bs n}_l(-L)\cdot{\bs n}(-L)$, where $\mathcal{T}$ is the coupling strength, while ${\bs n}_l(z)$ describes the magnetization of the lead whose dynamics is dictated by a similar Lagrangian as in Eq.~\eqref{lagrange} for $z\in(-\infty,-L)$. 

In the absence of an external magnetic field, the ferromagnet has two classical ground states, ${\bs n}(z)=\pm{\bs e}_z$, which are uniformly polarized along the easy axis.  Additionally, in the limit $L\rightarrow\infty$ the model above admits a DW solution ${\bs n}_0$ that minimizes the Hamiltonian part [second term in Eq.~\eqref{lagrange}] for boundary conditions ${\bs n}_0(z=\pm\infty)=\pm\bs{e}_z$: 
\begin{equation}
\cos\theta_0(z)=\tanh\left[Q(z-Z)\right];\,\phi(z)\equiv\Phi\,,
\end{equation}
where $Q=\int_{-\infty}^\infty dz\partial_z\theta/\pi=\pm1$ is the topological charge determined by the boundary conditions (at the edges of a long wire),  while $Z$ and $\Phi$ are the unconstrained position and angle of the domain wall, with all lengths expressed in terms of  $\lambda\equiv\sqrt{A/K_e}$. The low-energy dynamics of the DW for $b,K_h\neq 0$ is described by the Lagrangian \cite{TataraDW08}
\begin{align}
    \mathcal{L}_{dw}&=-2S_{dw}Z\dot{\Phi}+U(\Phi)-V_{\rm pin}(Z)\nonumber\,,\\
    U(\Phi)&=\widetilde{K}_h(\sin^2\Phi+\tilde{b}\sin\Phi)\,,
\end{align}
where $S_{dw}=s\lambda$, $\widetilde{K}_h=K_h\lambda$, $\tilde{b}=\pi b/2K_h$, and  we introduced also a pinning potential $V_{\rm pin}(Z)=-(\delta K_e/2)\cosh^{-2}(Z-Z_0)$, stemming from a point-like  defect at position $Z_0$ that causes a local reduction of the easy-axis anisotropy by $\delta K_e(>0)$ \cite{TataraDW08}. Using $P_{\Phi}=\partial_{\dot{\Phi}}\mathcal{L}_{dw}=-2S_{dw}Z$ and $P_{Z}=\partial_{\dot{Z}}\mathcal{L}_{dw}=0$, we can identify that $\{Z, \Phi\}$ are canonically conjugate variables. In the limit of strong pinning $\delta K_e\gg\widetilde{K}_h{\rm max}(1,\tilde{b})/S^2_{dw}$ (but $\delta K_e\ll K_e/s$),   $Z$ becomes a purely harmonic variable defining inertia of a generally nonlinear dynamics of $\Phi$ that is governed by Eq.~\eqref{DWham}.

In the presence of high-frequency fluctuations, the magnetization profile becomes ${\bs n}(z,t)={\bs n_0}(z-Z)+\delta{\bs n}(z,t)$, where $\delta{\bs n}(z,t)\perp{\bs n}_0(z-Z)$ describes the magnons. We can further write $\delta{\bs n}(z,t)={\bs e}_1\delta n_1(z,t)+{\bs e}_2\delta n_2(z,t)$ with ${\bs e}_1=\partial_{\theta_0}{\bs n}_0$ and ${\bs e}_2=\sin^{-1}\theta_0\partial_{\Phi}{\bs n}_0$, with $\delta n_{1,2}\ll1$ being the corresponding magnonic amplitudes. Let us assume that $K_h=b=0$, as well as $Z$ and $\Phi$ are fixed. Then, the resultant  Lagrangian describing the fluctuations is
\begin{align}
\mathcal{L}_{\rm m}=&-\frac{1}{2}\int_{-L}^L 
 dz[2s\delta n_1\partial_t\delta n_2+\delta n_1\mathcal{H}_0\delta n_1+\delta n_2\mathcal{H}_0\delta n_2]\,,\nonumber\\
\mathcal{H}_0&=\widetilde{K}_e\left(-\frac{\partial^2}{\partial z^2}+1-2{\rm sech}^2(z-Z)\right)\,,
\end{align}
with $\widetilde{K}_e=K_e\lambda$, and which describes magnons moving in a potential well $U(z) = 2{\rm sech}^2(z-Z)$. We can parameterize the magnonic field as $\delta n(z,t)\equiv \delta n_1(z,t)+i\delta n_2(z,t)=\sum_n\Psi_{n}(z,Z)a_n(t)$,
where $\Psi_{n}(z,Z)$ with $n=0,1,\dots$ are the standing wave solutions pertaining to a magnetic wire of length $2L$ described by the Hamiltonian $\mathcal{H}$, while $a_n(t)$ are the complex-valued amplitudes of the magnon in the state with energy $\omega_n$. In a quantum description, $a_n(t)$ [$a^*_n(t)$] becomes the magnon annihilation (creation) operator that satisfies $[a_n,a_m^\dagger]=\delta_{nm}$ \cite{SM}. For a finite wire of length $2L$, standing waves of energy $\omega_k$ can be written as $\Psi_k(z,Z)=\alpha_k\psi_k(z,Z)+\beta_k\psi_{-k}(z,Z)$, subject to the vanishing spin current at the boundaries $J_s^z\equiv\partial_z\Psi_k(\pm L,Z)=0$, while $\psi_k(z,Z)$ are normalized magnon solutions in an infinite system that can be found explicitly by exploiting the property that they scatter over the potential $U(z)$ without any reflection \cite{KimPRB18,SM}, resulting in
\begin{align}
    \psi_k(z,Z)=\frac{\tanh(z-Z)-ik}{\sqrt{s(1+k^2)}}e^{ikz}\,.
    \label{magnon_wf}
\end{align}
Above, $k$ is the magnon momentum along the wire, while the corresponding spectrum is $\omega_k=\omega_0(1+k^2)$, with $\omega_0=K_e/s$ being the magnonic gap.  When $L\gg1$ and the DW location is far from the edges, that is, $L-|Z|\gg1$, the boundary conditions result in the transcendental equation $\tan(2k_nL)=2k_n/(1-k_n^2)$ for the set of allowed magnonic momenta $k_n$, with $n=0,1,\dots$, while normalization of the wave function gives $\beta_n=\pm\alpha_n=1/{\sqrt{2L}}$ \cite{SM}. Here, the sign $+$ ($-$) corresponds to the set of solutions that satisfy $\tan(k_nL)=k_n$ [$\tan(k_nL)=-1/k_n$].  The energy separation between neighboring levels is $\Delta\omega_n\equiv\omega_{n+1}-\omega_{n}\approx\omega_0\pi^2(2n+1)/L^2$. Promoting $\Phi$ to a dynamical variable results in the following magnon-DW interaction in a frame rotating with the DW:
\begin{align}
\mathcal{L}_{int}(Z)&=\frac{s\dot{\Phi}}{2}\int_{-L}^L dz(\delta n_1^2+\delta n_2^2)\cos\theta_0\,,
\end{align}
which, as anticipated, pertains to a fictitious magnetic field that alters the DW Berry phase by the excited magnons 
\footnote{The long-wavelength magnons are practically unaffected by the pinning potential because their amplitude is negligible at the position of the scatterer,  according to Eq.~\eqref{magnon_wf} with $Z=Z_0$ (see also \cite{SM}).}.

{\it DW manipulation and detection}.|The dynamics of the wire induced by the Hamiltonian in Eq.~\eqref{DW_magnons} is captured by a Lindblad evolution of the density matrix $\rho_{\rm S}$ \cite{WilsonRae_NJP_08}:
\begin{align}
    \dot{\rho}_{\rm S}=i[\rho_{\rm S},H_{\rm tot}]&+\kappa_n\mathcal{D}(a_n)\rho_{\rm S}+\sum_{j,k}\gamma_{jk}\mathcal{D}(A_{jk})\rho_{\rm S}\,,
\label{DW_diss}    
\end{align}
where $\mathcal{D}(\mathcal{O})\rho_{\rm S}=\mathcal{O}\rho_{\rm S}\mathcal{O}^\dagger-\{\mathcal{O}^\dagger\mathcal{O},\rho_{\rm S}\}/2$ represents the dissipator, $\kappa_n$ is the total decay rate of the magnonic mode $n$ (including the intrinsic Gilbert damping), and $A_{jk}=|j\rangle\langle k|$. Above, we retained only the emission processes in the magnonic dissipator, which is justified in the limit $\omega_0\gg k_BT$ considered here. We can further expand the magnonic operators around their steady-state values, $a_n\rightarrow\alpha_n+a_n$, with $\alpha_n\approx-h_n/(\Delta_n-i\kappa_n/2)$, so that the resulting master equation is the same as above, but with the Hamiltonian:
\begin{align}
\!\!H'_{\rm tot}&\approx H_{\Phi}+g_{n}P_\Phi(a_n^\dagger+a_n)+\Delta_na_n^\dagger a_n\,,   
\end{align}
where $g_{n}=P_n(Z_0)|\alpha_n|/M_{\Phi}$, and the detunings $\Delta_n$ are slightly shifted by the DW-induced Lamb shifts of the magnons \cite{WilsonRae_NJP_08}. This Hamiltonian is identical to that describing (nonlinear) optomechanical setups \cite{MarquardtRMP14} and can therefore be used to manipulate the DW states.     

\begin{figure}[t]
\includegraphics[width=\linewidth]{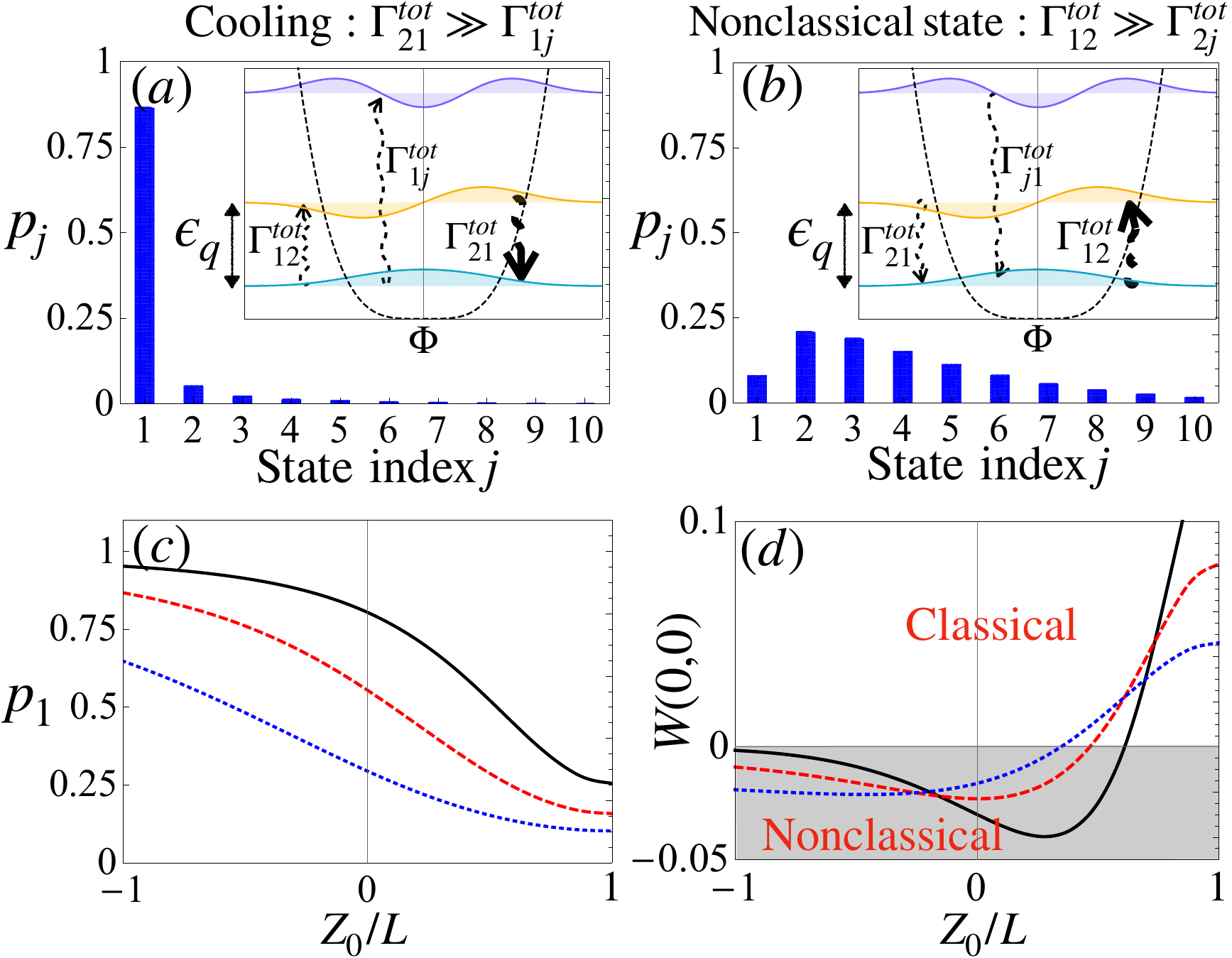}
\caption{(a) Occupation of the first ten DW levels when the uniform magnon ($n=1$) is driven such that $\Delta_{1}=-\epsilon_q$, and $Z_0=-L$. (b) Same, for $\Delta_{1}=\epsilon_q$ and $Z_0=0$, resulting in a nonclassical state with negative Wigner function, $W(0,0)=-0.02$. (c) The ground state population $p_1$ as a function of the position of the DW, $Z_0/L$, for $\Delta_{1}=-\epsilon_q$ (cooling). The black, dashed-red and dotted-blue curves correspond to $T=5\epsilon_q$,  $T=10\epsilon_q$, and $T=20\epsilon_q$, respectively. (d) The Wigner function $W(0,0)$ as a function of $Z_0/L$ for $\Delta_{1}=\epsilon_q$ (heating), with the same parameters as in (c); $W(0,0)<0$ (shaded area) describes nonclassical states. Note that $Z_0=1$ refers to a thermal distribution because the DW-magnon interaction vanishes. Also, $|\alpha_1|=0.1$, while all other parameters are the same as in Fig.~\ref{fig2}.}
\label{Fig3} 
\end{figure}

A first application concerns the sideband cooling of the DW levels via frequency-up conversion of the magnons, which could facilitate the initialization of the DW qubit. For efficient cooling, the following conditions must be met \cite{MarquardtRMP14} 
\begin{align}
(i)\,\, g_n|\langle 1|P_\Phi|2\rangle|\ll\kappa_n\ll\epsilon_q\,;\,\,\,(ii)\,\,\gamma_{21}\ll\frac{g_n^2|\langle 1|P_\Phi|2\rangle|^2}{\kappa_n}\,,\nonumber
\end{align}
allowing to eliminate adiabatically the magnons and ensuring that the cooling  overcomes the environmental heating of the DW via the Gilbert damping, respectively. The reduced density matrix that describes the dynamics of the DW, $\rho_\Phi={\rm Tr}_{\rm m}[\rho_{\rm S}]$,
obeys the Lindblad equation as in Eq. (10), but with bare Hamiltonian $H_{\Phi}$ and the
total transition rate from state $j$ to state $k$ given by  $\Gamma_{jk}^{\rm tot}=\gamma_{jk}+\Gamma^{n}_{jk}$, where
\begin{align}
    \Gamma^{n}_{jk}=\frac{1}{2}\frac{g_n^2|\langle j|P_\Phi|k\rangle|^2\kappa_n}{(\Delta_n+\epsilon_{jk})^2+(\kappa_n/2)^2}\,.
\end{align}
In the stationary limit, the DW density matrix becomes $\rho_\Phi=\sum_jp_j|j\rangle\langle j|$, where $p_j$ is the occupation of the level $j=1,2,\dots$. When $\Delta_n=-\epsilon_{q}$, the driven magnons can extract energy from the DW, cooling it towards the ground state, as depicted in Fig.~\ref{Fig3}(a). On the other hand, when $\Delta_n=\epsilon_{q}$ the interplay of the driven magnons and the anharmonicity of the DW spectrum results in an enhancement of $p_{2}$ with respect to the other DW levels, as shown in Fig.~\ref{Fig3}(b) for $n=1$ (uniform magnonic mode). Importantly, such a state can exhibit a negative Wigner function at the origin $W(\Phi=0,P_\Phi=0)=\sum_{j}(-1)^{j+1}p_j$ \cite{Rips_NJP_2012,Harochebook} that characterizes a nonclassical DW state (see also \cite{SM}).  

The steady state of a DW can be probed
with an additional microwave field, weakly driving one magnonic mode $\omega_p$ on resonance, i.e. with $\Omega_p=\omega_p$ (e.g. the $p=2$ magnonic mode). The power spectrum of the emitted magnons reads 
\begin{align}
    \mathcal{S}_p(\omega)=\frac{1}{2\pi}\int_{-\infty}^\infty d\tau e^{-i(\omega-\Omega_p)\tau}\langle a^\dagger_{p,\rm out}(t+\tau)a_{p,\rm out}(t)\rangle\nonumber\,,
\end{align}
where the output field $a_{p,\rm out}(t)$ is related to the magnonic mode $a_p$ via the standard input-output relation $a_{p,\rm out}(t)=\sqrt{2\kappa_{p,ex}}a_p(t)+a_{p,\rm in}(t)$, and $\langle\dots\rangle$ means average over the stationary state of the DW. The dynamics of the intra-cavity
field can be described by a quantum Langevin equation \cite{RaePRA12}:
\begin{align}
    \dot{a}_p(t)&=-\frac{\kappa_p}{2}a_p(t)+i\frac{P_p(Z_0)P_\Phi}{M_\Phi} a_p(t)+\sqrt{\kappa_{p,ex}}a_{p,\rm in}(t)\nonumber\,,
    \label{in_out}
\end{align}
where $a_{p,\rm in}(t)$ represents the input field in the probe mode, $\kappa_{p,ex}$ the decay rate of mode $p$ into the leads,  and all operators are in the Heisenberg picture. Eq.~\eqref{in_out} can be formally integrated, and then subjected to a Dyson series-type expansion to the first order in $g_p$.  Assuming $\Gamma_{jk}^p\ll \Gamma^{\rm tot}_{jk}$, such that the measurement does not affect the DW level occupations induced by the other drivings, we find for the side-band transitions (i.e. $\omega\neq\omega_p$) \cite{SM}:
\begin{align}
\mathcal{S}_p(\omega)\approx\frac{\kappa_{p,ex}}{\pi\kappa_p}\sum_{j,k}\frac{\Gamma^p_{jk}\Gamma^{\rm eff}_{jk}}{(\omega-\Omega_p-\epsilon_{jk})^2+(\Gamma^{\rm eff}_{jk})^2}p_j\,,
\end{align}
where $\Gamma^p_{jk}=\Gamma^p_{kj}$ at $\Delta_p=0$, while $\Gamma_{jk}^{\rm eff}=\sum_s(\Gamma_{sk}^{\rm tot}+\Gamma_{sj}^{\rm tot})=\Gamma_{kj}^{\rm eff}$ represents the peak's linewidth. Consequently, the power spectrum of the emitted magnons exhibits a side-band structure because of the DW-assisted frequency up- or down-conversion processes. For example, when $\Delta_n=-\epsilon_{q}$ (ground-state cooling), $\mathcal{S}_p(\Omega_p-\epsilon_{q})/\mathcal{S}_p(\Omega_p+\epsilon_{q})\approx p_1/p_2$, which allows us to infer the ground-state population from the power spectrum of the emitted magnons (together with $\sum_jp_j=1$). 

Magnons can mediate interactions between two such DWs, facilitating the implementation of two-qubit gates. Indeed, assuming that both DWs are cooled down to their respective ground states, and sufficient nonlinearity in the spectrum, their dynamics are restricted to the subspace spanned by the lowest two states, $\{|1_{a}\rangle,|2_{a}\rangle\}$, with $a=1,2$ labeling the DW. In the limit $g_{n,a}\ll||\Delta_{n}|-\epsilon_{q,a}|\ll||\Delta_{n}|+\epsilon_{q,a}|$ (dispersive regime), we can eliminate perturbatively the magnons (here neglecting their decay $\kappa_n$) to obtain $H_{1-2}=J_{1-2}(\sigma_1^+\sigma_2^-+{\rm h. c. })$ with the exchange coupling
\begin{align}
    J_{1-2}=\sum_{a=1,2}\frac{g_{n,1}g_{n,2}|\langle1_1|P_{1,\Phi}|2_1\rangle||\langle1_2|P_{2,\Phi}|2_2\rangle|}{\Delta_{n}-\epsilon_{q,a}}\,,
\end{align}
where we assumed identical DWs, and $\sigma_a^+=|2_a\rangle\langle1_a|$ ($\sigma_a^-=|1_a\rangle\langle2_a|$) are the raising (lowering) Pauli operators associated with qubit $a=1,2$. Above, we also performed the rotating wave approximation (therefore neglecting terms $\propto\sigma_1^+\sigma_2^+,\sigma_1^-\sigma_2^-$) and disregarded small corrections to each individual qubit Hamiltonian \cite{RipsPRL12}. The evolution of the combined system with $H_{1-2}$ for $t_g=\pi/4J_{1-2}$, implements a $\sqrt{i{\rm SWAP}}$ operation which, up to single-qubit rotations, is equivalent to a controlled-NOT gate \cite{BarencoPRA95}. Together with single-qubit gates, which can be implemented by locally pulsing $ac$ magnetic fields, the interaction $H_{1-2}$ is therefore sufficient for universal quantum computation.

{\it Conclusions}.|In summary, we have shown that magnons in short insulating ferromagnetic wires can be leveraged to initialize, entangle and detect quantum states encoded in macroscopic DWs. These features are enabled by universal (and intrinsic) Berry-phase mediated interactions between the magnons and the DW,  rendering our approach promising in terms of scalability, since it does not require extrinsic degrees of freedom (e.g. photons) to be integrated with the ferromagnetic wire. By the same token, it could be naturally extended to 2D topological objects, such as skyrmions, or even to other magnetic materials, such as ferrimagnets or antiferromagnets, which can exhibit faster time scales for manipulations.

The authors acknowledge fruitful discussions with Dalton Jones in the early stages of this work. This work is supported by the Foundation for Polish Science through the international research agendas program co-financed by the European Union within the smart growth operational program,  by the National Science Centre (Poland) OPUS 2021/41/B/ST3/04475 (MT) and NSF under Grant No. DMR-2049979 (YT).

%



\onecolumngrid
\clearpage

\setcounter{secnumdepth}{3}
\makeatletter
\xdef\presupfigures{\arabic{figure}}
\setcounter{equation}{0}
\setcounter{table}{0}
\setcounter{page}{1}
\renewcommand{\thefigure}{S\fpeval{\arabic{figure}-\presupfigures}}
\renewcommand{\theequation}{S\@arabic\c@equation}
\renewcommand{\thetable}{S\@arabic\c@table}
\renewcommand{\thepage}{S\@arabic\c@page}
\begin{widetext}
	
	\newpage
	\onecolumngrid
	\bigskip 
	
	\begin{center}
		\large{\bf Supplementary Material for ``Cavity magnonics with domain walls in insulating ferromagnetic wires" \\}
	\end{center}
	\begin{center}
		Mircea Trif$^1$ and Yaroslav Tserkovnyak$^{2,3}$
		\\
\it{$^1$International  Research  Centre  MagTop,  Institute  of  Physics,  Polish  Academy  of  Sciences, Aleja  Lotnikow  32/46,  PL-02668  Warsaw,  Poland}\\
\it{$^2$Department of Physics and Astronomy, University of California, Los Angeles, USA}\\
\it{$^3$Bhaumik Institute for Theoretical Physics, University of California, Los Angeles, CA 90095, USA}
\end{center}
\date{\today}

\section{Isolated DW dynamics}

As described in the main text, the isolated dynamics of a DW positioned far from the boundaries, i.e. $L\gg Z$, is governed by the Lagrangian \cite{TataraDW08}:   
\begin{align}
    \mathcal{L}_{dw}\approx-2S_{dw}Z\dot{\Phi}-\widetilde{K}_h(\sin^2\Phi+\tilde{b}\sin\Phi)-V_{\rm pin}(Z)\,,
\label{L_dw_SM}
\end{align}
where $S_{dw}=s\lambda$, $\widetilde{K}_h=K_h\lambda$, and $\tilde{b}=\pi b/2K_h$.  Furthermore, the pinning potential $V_{\rm pin}(Z)$ (around $Z=Z_0$) may be due to inhomogeneous magnetic anisotropies in the sample (e.g., reduced local $K_e$ that can be controlled electrically). The dissipation function, on the other hand, becomes
\begin{align}
    \mathcal{R}_{dw}=\frac{S_{dw}\alpha}{2}\left(\dot{Z}^2+\dot{\Phi}^2\right)\,.
\end{align}
The equations of motion engendered by the Lagrangian stated in Eq.~\eqref{L_dw_SM} are
\begin{align}
P_{\Phi}&=\frac{\partial\mathcal{L}_{dw}}{\partial\dot{\Phi}}=-2S_{dw}Z\,,\,\,\,\,P_{Z}=\frac{\partial \mathcal{L}_{dw}}{\partial\dot{Z}}=0\,,
\end{align} 
which indicates that  $\{Z, \Phi\}$ are Hamiltonian canonically conjugate variables. With dissipation, the generalized Euler-Lagrange equations of motion read:
 \begin{align}
\widetilde{K}_h[\sin{(2\Phi)}+\tilde{b}\cos{\Phi}]&-2S_{dw}\dot{Z}=-\alpha S_{dw} \dot{\Phi}\,,\nonumber\\
2S_{dw}\dot{\Phi}+\frac{\partial V_{\rm pin}(Z)}{\partial Z}&=-\alpha S_{dw}\dot{Z}\,.
 \end{align} 
By replacing $\dot{Z}$ in the second line and taking the time-derivative of the equation that follows, we can obtain (while only considering the Rayleigh dissipation of $\Phi$ in the case of strong pinning): 
\begin{align}
2S_{dw}\ddot{\Phi}&+\frac{\partial^2 V_{\rm pin}(Z)}{\partial Z^2}\left(\widetilde{K}_h(\sin{(2\Phi)}+\tilde{b}\cos{\Phi})+\alpha S_{dw}\dot{\Phi}\right)/2S_{dw}=0\,,\nonumber\\
M_\Phi\ddot{\Phi}&+\widetilde{K}_h\left(\sin(2\Phi)+\tilde{b}\cos\Phi\right)+\alpha S_{dw}\dot{\Phi}=0\,, 
\end{align}
where  
\begin{equation}
    M_\Phi=(2S_{dw})^2\left(\frac{\partial^2 V_{\rm pin}(Z)}{\partial Z^2}\right)^{-1}
\end{equation}
is the inertia of the DW associated with rigid rotations. Assuming
\begin{align}
    V_{\rm pin}(Z)\approx\frac{\delta K_e(Z-Z_0)^2}{2}\,,
\end{align}
results in the mass $M_\Phi=(2S_{dw})^2/\delta K_e$. The effective Lagrangian describing the dynamics of the angle $\Phi$ only (in the absence of dissipation) becomes
\begin{equation}
\mathcal{L}_{\Phi}=\frac{1}{2}M_{\Phi}\dot{\Phi}^2-\underbrace{\widetilde{K}_h(\sin^2\Phi+\tilde{b}\sin\Phi)}_{\displaystyle U(\Phi)}\,,
\end{equation}
which in turn pertains to the following effective Hamiltonian 
\begin{align}
H_{\Phi}&=\frac{1}{2M_{\Phi}}P_{\Phi}^2+U(\Phi)\,,
\end{align}
which is depicted in Eq.~(1) in the main text. 

We can now promote the momentum $P_{\Phi}$ to its operator version, $P_{\Phi}\rightarrow-i\partial_\Phi$, which eventually allows determining the Hamiltonian that describes the quantum dynamics of the isolated DW. 

To make progress on that end, it is beneficial to identify the classical potential minima of $U(\Phi)$, which implicate solving
\begin{align}
\cos\Phi_{eq}\left(\sin\Phi_{eq}-\tilde{b}\right)=0\,, 
\end{align}
along with the local maximum condition
\begin{equation}    2\cos2\Phi_{eq}+\tilde{b}\sin\Phi_{eq}\geq0\,.
\end{equation}
We see that $\Phi_{eq}=\pi/2$ becomes an inflection point at $\tilde{b}_{c}=1$ (rendering the effective potential quartic in $\Phi$), while the two minima before reaching this point are located at $\{\Phi_{eq}^1,\Phi_{eq}^{2}\}=\{-\pi+\arcsin{\tilde{b}},-\arcsin{\tilde{b}}\}$. 

There are three regimes for the operation of the DW: ($i$) $0<\tilde{b}<1$, ($ii$) $\tilde{b}=1$ (the case studied in detail in this work), and ($iii$) $\tilde{b}>1$.

\subsubsection{Double-well potential limit: $0\leq\tilde{b}<1$}

The first case corresponds to a double-well potential and has been discussed in many previous works (see, for example,~\cite{zou2022domain}). To determine the spitting of the lowest levels, we calculate the amplitude of the transition between the two minima, $G(\Phi_{eq}^1,\Phi_{eq}^2)$, using the Euclidian path integral approach by switching to the imaginary time representation $t\rightarrow i\tau$. Thus, we write:
\begin{align}
G(\Phi_{eq}^1,\Phi_{eq}^2,\tau)&\equiv\langle\Phi_{eq}^1|e^{-\tau H_{\Phi}}|\Phi_{eq}^2\rangle=\int_{\Phi(0)=\Phi_{eq}^1,\Phi(\tau)=\Phi_{eq}^2}{\mathcal D}\Phi\,e^{-\int_0^\tau d\tau'{\mathcal L}_{\Phi}(\Phi,\partial_{\tau'}\Phi)}\,,
\end{align}
where
\begin{equation}
{\mathcal L}_{\Phi}(\Phi,\partial_{\tau'}\Phi)=\frac{1}{2}M_{\Phi}\dot{\Phi}^2+U(\Phi)
\end{equation}
represent the Lagrangian for the DW particle moving in the inverted potential $-U(\Phi)$. 
The corresponding classical equations of motions hence read as follows
\begin{align}
M_{\Phi}\ddot{\Phi}_{cl}-\frac{\partial U(\Phi_{cl})}{\partial\Phi_{cl}}=0\,,
\end{align}
so that the classical trajectories $\Phi_{cl}$ will satisfy:
\begin{equation}
\frac{1}{2}M_{\Phi}\dot{\Phi}_{cl}^2-U(\Phi_{cl})={\rm const}\,.
\end{equation}
We are looking for solutions that do not allow the particle to ``escape'' the maxima of the inverted potential, that is, $\dot{\Phi}|_{\Phi_{eq}^{1,2}}=0$, so we get const$=U(\Phi_{eq}^{1,2})\equiv U_c$. We can define a new (shifted) potential $U'(\Phi_{cl})=U(\Phi_{cl})-U_c$ such that $U'(\Phi_{eq}^{1,2})=0$ (note that adding a constant term does not influence the dynamics). Thus, the classical trajectory Lagrangian is given by ${\mathcal L}_{\Phi}(\Phi_{cl},\partial_{\tau}\Phi_{cl})=M_\Phi\dot{\Phi}_{cl}^2\equiv\sqrt{2M_\Phi U'(\Phi_{cl})}\dot{\Phi}_{cl}$. Since the lowest barrier between the two minima is $\Phi=-\pi/2$, we can retain in the action only the trajectories that involve this barrier. The classical action is then obtained:
\begin{align}
{\mathcal S}_{cl}&=\int_0^\tau d\tau'{\mathcal L}_{\Phi}(\Phi_{cl},\partial_{\tau}\Phi_{cl})=\int_{\Phi_{eq}^1}^{\Phi_{eq}^2} d\Phi_{cl}\sqrt{2M_\Phi U'(\Phi_{cl})}=2\sqrt{2M_{\Phi}}\left[\sqrt{1-\tilde{b}^2}-\tilde{b}\arccos{\tilde{b}}\right]\,.
\end{align}  
Following the discussion in~\cite{AltlandBook}, we find the following.
\begin{align}
G(\Phi_{eq}^1,\Phi_{eq}^2,\tau)&\approx Ce^{-\omega\tau}\sum_{n\in{\rm odd}}\frac{(\tau K)^{n}}{n!}e^{-n\mathcal{S}_{cl}}=Ce^{-\omega\tau}\sinh{\left[\tau Ke^{-\mathcal{S}_{cl}}\right]}\,,
\end{align}
where $\omega$ is the oscillation frequency associated with one of the wells only
\begin{equation}
    \omega=\sqrt{\frac{2\widetilde{K}_h(1-\tilde{b}^2)}{M_\Phi}}=\frac{1}{S_{dw}}\sqrt{\frac{\delta K_e\widetilde{K}_h(1-\tilde{b}^2)}{2}}\,,
\end{equation}
while $C$ is a constant that can be extracted from treating an isolated well analytically \cite{AltlandBook}. Moreover,
\begin{equation}
K=\gamma\omega\sqrt{\frac{\mathcal{S}_{cl}}{2\pi}}\,,
\end{equation}
where $\gamma\approx1$. Physically, the two initially degenerate minima at $\Phi_{eq}^{1,2}$ become mixed due to macroscopic quantum tunneling, and the states split into symmetric $|S\rangle$ and antisymmetric $|A\rangle$ configurations of the wave functions. Projecting the transition probability onto this subspace, we can write:
\begin{align}
G(\Phi_{eq}^1,\Phi_{eq}^2,\tau)&\approx\langle\Phi_{eq}^1|\left(|S\rangle e^{-\epsilon_S\tau}\langle S|+|A\rangle e^{-\epsilon_A\tau}\langle A|\right)|\Phi_{eq}^2\rangle\nonumber\\
&=\frac{C}{2}e^{-\bar{\epsilon}\tau/}\left(e^{-\Delta\epsilon\tau/2}-e^{\Delta\epsilon\tau/2}\right)=Ce^{-\bar{\epsilon}\tau}\sinh{(\tau\Delta\epsilon)}\,,
\end{align} 
with $\bar{\epsilon}=(\epsilon_A+\epsilon_S)/2$, $\Delta\epsilon=\epsilon_A-\epsilon_S$, and $C=2|\langle\Phi_{eq}^{1,2}|S\rangle|^2$. Comparing the above expression with the quasi-classical calculation, we obtain the following.
\begin{align}
\epsilon_{S,A}=\frac{\omega}{2}\pm Ke^{-\mathcal{S}_{cl}}=\frac{\omega}{2}\pm\gamma\omega\sqrt{\frac{\mathcal{S}_{cl}}{2\pi}}e^{-\mathcal{S}_{cl}}\,,
\end{align} 
and thus the splitting between the two states (qubit splitting) is:
\begin{equation}
\Delta\epsilon=2\gamma\omega\sqrt{\frac{\mathcal{S}_{cl}}{2\pi}}e^{-\mathcal{S}_{cl}}\,.
\end{equation}
Above, we assumed $\langle\Phi_{eq}^1|\Phi_{eq}^2\rangle\approx0$, i.e. the two states localized at the potential minima have negligible overlap (such a condition only applies for large barriers).

\subsubsection{Single-well potential limit: $\tilde{b}\geq1$}

In this case, $U(\Phi)$ can be expanded around the minimum located at $\Phi_0=3\pi/2$, which will be set to zero from now on. Hence, writing $\Phi\rightarrow\Phi_0-\Phi$, we obtain:
\begin{align}
    U(\Phi)\approx \underbrace{\widetilde{K}_h(\tilde{b}-1)}_{A}\Phi^2+\underbrace{\frac{\widetilde{K}_h(4-\tilde{b})}{12}}_{B}\Phi^4\,,
\end{align}
near the critical point $\tilde{b}\geq1$.  

First, we establish the spectrum at the transition point $\tilde{b}=1$. In that case, $A=0$, and $B=\widetilde{K}_h/4$. Although this can be solved exactly only numerically, we can employ the quasi-classical (Bohr-Sommerfeld quantization) approach to estimate the eigenenergies and the corresponding level spacing. That is:
\begin{align}
    \int_{-\Phi_j}^{\Phi_j}d\Phi\sqrt{2M_{\Phi}[\epsilon_j-U(\Phi)]}=\left(j-\frac{1}{2}\right)\pi\,,
\end{align}
where $j=1,2,3,\dots$, and $\pm \Phi_j=\pm(\epsilon_j/B)^{1/4}$ represent the classical turning points. We find:
\begin{align}
    \epsilon_j&=\left[\frac{\pi}{\mathcal{C}}\frac{B^{1/4}}{\sqrt{2M_{\Phi}}}\right]^{4/3}\left(j-\frac{1}{2}\right)^{4/3}\,,\nonumber\\
    \mathcal{C}&=\int_{-1}^1dx\sqrt{1-x^4}\approx 1.75\,.
\end{align}
Therefore, we can write $\epsilon_j=\epsilon_1(j-1/2)^{4/3}$, with
\begin{align}
    \epsilon_1\approx\frac{0.86}{S_{dw}^{4/3}}\left(\widetilde{K}_h\delta K_e^2\right)^{1/3}\,.
\end{align}
We see that the spectrum is highly anharmonic $\epsilon_2-\epsilon_1\approx1.32\epsilon_1$, $\epsilon_3-\epsilon_2\approx1.67\epsilon_1$, which is beneficial for any quantum information processing with these states.

\begin{figure}[t]
\includegraphics[width=\linewidth]{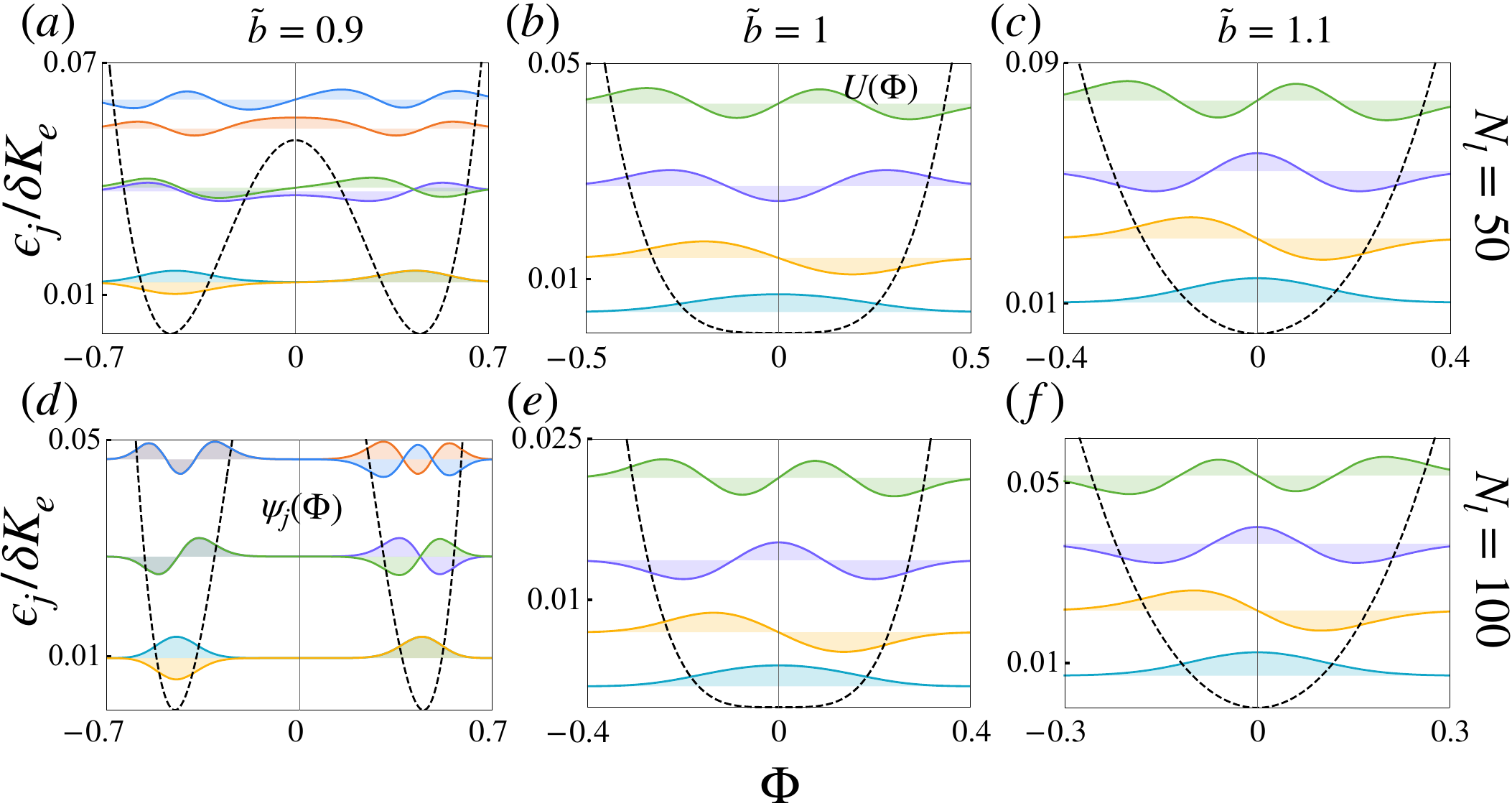}
\caption{The DW energy levels $\epsilon_j$ (scaled with $\delta K_e$), and the wave-functions $\psi_j(\Phi)$, with $j=1,2,3,\dots$ for several reduced magnetic fields 
 around $\tilde{b}=\pi b/2K_h=1$ that control the effective potential $U(\Phi)$ (dashed line). (a)--(c) correspond to $\widetilde{K}_h=5\delta K_e$, $S_{dw}=50$, and $N_l\equiv\lambda/a=50$ unit cells along the wire, while (d)--(f) $\widetilde{K}_h=10\delta K_e$ and $S_{dw}=100$, and $N_l=100$ unit cells along the wire. }
\label{fig1SM} 
\end{figure}

Next, let us focus on the $\tilde{b}>1$ regime.  In this case, the quadratic term is non-zero, and we can define the (harmonic) frequency associated with it as:  
\begin{align}
    \Omega=\sqrt{\frac{2A}{M_\Phi}}=\frac{1}{S_{dw}}\sqrt{\frac{(\tilde{b}-1)\widetilde{K}_h\delta K_e}{2}}\,,
\end{align}
as well as introducing:
\begin{align}
    \Phi=\sqrt{\frac{1}{2M_{\Phi}\Omega}}(a^\dagger+a)\,, P_\Phi=i\sqrt{\frac{M_{\Phi}\Omega}{2}}(a^\dagger-a)\,,
\end{align}
where $a$ ($a^\dagger$) is the annihilation (creation) operator associated with the harmonic motion. 
That in turn gives:
\begin{align}
    H_\Phi&=\hbar\Omega a^\dagger a+\frac{\Lambda}{2}(a^\dagger+a)^4\,,\nonumber\\
    \frac{\Lambda}{2}&=\frac{B}{(2M_{\Phi}\Omega)^2}=\frac{(4-\tilde{b})\delta K_e}{96S_{dw}^2(\tilde{b}-1)}\,.
\end{align}
This can be further simplified by performing the rotating wave approximation (RWA), leading to
\begin{align}
    H_\Phi=\Omega a^\dagger a+\frac{\Lambda}{2}\left[(a^\dagger)^2a^2+a^2(a^\dagger)^2+a^\dagger aa^\dagger a+aa^\dagger a a^\dagger+a^\dagger a aa^\dagger+aa^\dagger a^\dagger a\right]\nonumber
    &=\Omega'a^\dagger a+\frac{\Lambda'}{2}a^\dagger a^\dagger aa\,,
\end{align}
where $\Omega'=\Omega+\Lambda'$ and $\Lambda'=6\Lambda$. Here, the
RWA is warranted, provided that the relevant DW localized magnon numbers $n=0,1,2,\dots$, satisfy $n^2\Lambda\ll\Omega$. 
Finally, let us estimate the maximal number of photons under which the above assumption works:
\begin{align}
    n^2<n^2_{max}=\frac{\Omega}{\Lambda}=24S_{dw}\frac{(\tilde{b}-1)^{3/2}}{4-\tilde{b}}\sqrt{\frac{\widetilde{K}_h}{\delta K_e}}\,.
\end{align}
Assuming $S_{dw}=50$, $\tilde{b}=1.5$,  and $\widetilde{K}_h=5\delta K_e$, we find $n_{max}\approx20$.

A comparison between all regimes is shown in Fig.~\ref{fig1SM}, where the spectrum and wave functions have been found using full diagonalization.

\subsection{Dissipation via Gilbert damping}

Here, we discuss briefly how to account for the dissipation introduced by the Gilbert damping in the quantized Hamiltonian. We start by writing the classical equation of motion pertaining to the DW Lagrangian in the presence of the damping ($\mathcal{R}_{\Phi}=\alpha S_{dw}\dot{\Phi}^2/2$):   
\begin{align}
M_\Phi\ddot{\Phi}+\widetilde{K}_h[\sin(2\Phi)+\tilde{b}\cos\Phi]+\alpha S_{dw}\dot{\Phi}=\xi(t)\,, 
\end{align}
where $\xi(t)$ is the Langevin noise that is subject to the fluctuation-dissipation condition \cite{TakeiPRB17}
\begin{align}
\int_{-\infty}^\infty dt e^{-i\omega t}\langle\xi(t)\xi(0)\rangle=\alpha\omega S_{dw}\coth(\omega/2k_BT)\,.
\end{align}
In the quantum Hamiltonian description, this in turn can be included as:
\begin{align}
H_{\Phi}\rightarrow H_{\Phi}-\Phi\hat{\xi}+H_\xi\,,
\end{align}
where $\hat{\xi}=\sum_pg_p(\xi_p+\xi_p^\dagger)$ describe the bosonic environment with $\xi_p$ ($\xi_p^\dagger$) being the annihilation (creation) operators satisfying $[\xi_p,\xi_{p'}^\dagger]=\delta_{pp'}$, while $H_\xi$ dictates the environment dynamics. Following the approach described in one of the following sections, we can write the evolution of the DW density matrix as follows:
\begin{align}
\dot{\rho}_\Phi=-i[H_\Phi,\rho_\Phi]+\sum_{j,k}\gamma_{jk}\mathcal{D}(A_{jk})\rho_\Phi\,,
\end{align}
where $A_{jk}=|k\rangle\langle j|$, $\mathcal{D}(\mathcal{O})\rho_\Phi=\mathcal{O}^\dagger\rho_\Phi \mathcal{O}-\{\mathcal{O}^\dagger \mathcal{O},\rho_\Phi\}/2$ is the dissipator, while
\begin{align}
    \gamma_{jk}=2\alpha S_{dw}|\langle j|\Phi|k\rangle|^2(\epsilon_j-\epsilon_k)[1+n_B(\epsilon_j-\epsilon_k)]=\gamma_{kj}e^{(\epsilon_j-\epsilon_k)/k_BT}\,,
\end{align}
are the resulting rates that satisfy the detailed balance condition in thermal equilibrium. Above, 
\begin{align}
n_B(\omega)=\frac{1}{e^{\omega/k_BT}-1}\,,
\end{align}
is the Bose-Einstein distribution.  

\section{Description of the magnons}

To address the magnons' dynamics and their coupling to the DW, we first analyze the case when the latter is static. This will allow us to express the dynamical solutions in terms of the instantaneous basis for a given position $Z$ and orientation $\Phi$ of the DW.   

\subsection{Magnons around a static DW at $K_h=b=0$}

The fluctuations around the DW magnetization texture ${\bs n}_0(z)=(\sin\theta_0\cos\Phi, \sin\theta_0\sin\Phi, \cos\theta_0)\equiv {\bs n}_0(z-Z)$ can be described as follows:
\begin{align}
    {\bs n}(z,t)&=\frac{1}{\sqrt{1+\delta n_1^2+\delta n_2^2}}({\bs n}_0+{\bs e}_1\delta n_1+{\bs e}_2\delta n_2)\approx\left(1-\frac{\delta n_1^2+\delta n_2^2}{2}\right){\bs n}_0(z)+{\bs e}_1\delta n_1+{\bs e}_2\delta n_2\,,
    \label{exp_mag}
\end{align}
where 
\begin{align}
    {\bs e}_1&=\frac{\partial{\bs n}_0}{\partial\theta_0}=(\cos\theta_0\cos\Phi,\cos\theta_0\sin\Phi,-\sin\theta_0);\,\,{\bs e}_2=\frac{\partial{\bs n}_0}{\sin\theta_0\partial\Phi}=(-\sin\Phi,\cos\Phi,0)
\end{align}
are the unit vectors that are perpendicular to the DW magnetization at position $z$ (and mutually perpendicular), while $\delta n_{1,2}\equiv\delta n_{1,2}(z,t)$ are their amplitudes with respect to the DW position $Z$. Note that in the regime $\delta n_{1,2}(z,t)\ll1$, we have retained only terms up to the second order in the fluctuations.

The resulting instantaneous magnon Lagrangian becomes:
\begin{align}
\mathcal{L}_{sw}&=-\frac{1}{2}\int dz[2S_{dw}\delta n_1\partial_t\delta n_2+\delta n_1\mathcal{H}_0\delta n_1+\delta n_2\mathcal{H}_0\delta n_2]\,,\nonumber\\
\mathcal{H}_0&=\widetilde{K}_e\left(-\frac{\partial^2}{\partial z^2}+1-2{\rm sech}^2(z-Z)\right)\,,
\end{align}
which means magnons moving in a P\"{o}sch and Teller potential well $U(z)=-2{\rm sech}^2(z-Z)$, having the property that in an infinite system, the magnons pass through it without any reflection. In the following, we first determine the magnon solutions and spectrum for an infinite wire and then utilize these solutions to find the eigenmodes in the finite wire. We can formally write:
\begin{align}
\mathcal{H}_0&=\widetilde{K}_e\left(-\frac{\partial^2}{\partial z^2}+1-2{\rm sech}^2(z-Z)\right)=\widetilde{K}_ea^\dagger a\,,
\end{align}
with 
\begin{equation}
a\equiv \frac{\partial}{\partial z}+\tanh{(z-Z)},\,\,\,\,a^\dagger\equiv-\frac{\partial}{\partial z}+\tanh{(z-Z)}\,.
\end{equation} 
The Euler-Lagrange equations in the absence of damping lead to the following equations of motion for the fields $\delta n_{1,2}$:
\begin{align}
i s\left(
\begin{array}{c}
\delta \dot{n}_1\\
\delta \dot{n}_2
\end{array}
\right)
=\left(
\begin{array}{cc}
0 & i\mathcal{H}_0\\
-i\mathcal{H}_0 & 0
\end{array}
\right)
\left(
\begin{array}{c}
\delta n_1\\
\delta n_2
\end{array}
\right)\,,
\end{align}
which can be solved efficiently by using the techniques of super-symmetric quantum mechanics \cite{KimPRB18}. We briefly present the rationale behind this approach. Let us first define another Hamiltonian density:
\begin{align}
\mathcal{H}_0'&=\widetilde{K}_e\left(-\frac{\partial^2}{\partial z^2}+1\right)=\widetilde{K}_eaa^\dagger\,,
\end{align}  
which is translation invariant and allow for plane wave solutions $\delta n_0(z,t)\propto e^{i(kz-i\omega t)}$. The solutions to the initial problem are then simply found from $\psi(z)=a^\dagger\psi_0(z)$, while the two Hamiltonians share the same spectrum $\omega_k$. Let us thus solve the simpler problem:
\begin{align}
\omega\left(
\begin{array}{c}
\delta n_1^0(\omega,k)\\
\delta n_2^0(\omega,k)
\end{array}
\right)
=\frac{K_e(k^2+1)}{s}\sigma_y
\left(
\begin{array}{c}
\delta n_1^0(\omega,k)\\
\delta n_2^0(\omega,k)
\end{array}
\right)\,,
\end{align}
where $\sigma_y$ is a Pauli matrix acting in the space of the two orthogonal fluctuations. By applying the operator $\sigma_y$ again on the right, we obtain the magnons' spectrum and the two types of eigenvectors:
\begin{align}
\omega_k&=\pm\frac{K_e}{s}(k^2+1)\,,\nonumber\\
\Psi_{\pm}(k)&=\frac{1}{\sqrt{2}}\left(
\begin{array}{c}
1\\
\pm i
\end{array}
\right)\,,
\end{align} 
with $\pm$ corresponding to $\pm\omega_k$. While in a classical theory we can choose either sign of the frequency, the quantum description requires more caution. Let us calculate the energy of the mode:
\begin{align}
    \frac{1}{2}(\delta n_{1,s}^0H_0\delta n_{1,s}^0+\delta n_{2,s}^0H_0\delta n_{2,s}^0)\propto\frac{1}{2}\Psi_{s}^\dagger H_0\Psi_s=\omega_k\,, 
\end{align}
and therefore each of the negative and positive eigenvalue solution correspond to an energy $|\omega_k|$. Therefore, the fields $\delta n_{1,2}^0(z,t)$ can be presented in a compact way as: 
\begin{align}
    \delta{\bs n}^0(z,t)=[\delta n_1^0(z,t),\delta n_2^0(z,t)]^T=\frac{1}{\sqrt{2}}\sum_{k,s=\pm}a_{k,\pm}(t)\Psi_{\pm}\psi_k^0(z)e^{\pm i\omega_kt}\,,
\end{align}
and, imposing the condition that the fields are real, we find:
\begin{align}
a_{k,\pm}=(a_{-k,\mp})^*\equiv a_k\,.    
\end{align}
We can conclude that $\delta n_1^0(z,t)$  and $\delta n_2^0(z,t)$ are related to the Re and Im parts of the same function $\delta n^0(z,t)=\delta n_1^0(z,t)+i\delta n_2^0(z,t)$. Therefore, we can write
\begin{align}
\delta n^0(z,t)&\equiv\sum_ka_k(t)\psi_k^0(z)e^{i\omega_kt}
\,,
\end{align}
where 
\begin{equation}
a_k(t)=\int dz \delta n^0(z,t)(\psi^0_k)^*(z)\,,
\end{equation}
is the amplitude of the spin wave with momentum $k$ that oscillates with frequency $\sim\omega_k$, and $\psi_k^0(z)=e^{ikz}$ is the associated wave-function. This in turn allows us to find the actual fields $\psi_k(z)=a^\dagger\psi_k^0(z)$, that is,
\begin{equation}
\psi_k(z,Z)=\frac{\tanh(z-Z)-ik}{\sqrt{s(1+k^2)}}e^{ikz}\equiv\psi^*_{-k}(z)\,.
\end{equation} 
The factor $1/\sqrt{s(1+k^2)}$ was chosen to ensure that the following normalization condition is maintained:
\begin{equation}
s\int_{-L}^L dz\psi^*_k(z)\psi_{k'}(z)=\frac{1}{\sqrt{(1+k^2)(1+k'^2)}}\int_{-L}^L dz\left[e^{i(k-k')z}(1+kk')-\frac{d}{dz}\left(e^{i(k-k')z}\tan(z-Z)\right)\right]=2\pi\delta(k-k')\,,
\end{equation}
where we neglected the last term that gives a highly oscillatory contribution when $L\rightarrow\infty$. Finally, a general solution can be written
as follows
\begin{align}
    \delta n(z,t)=\sum_ka_k(t)\psi_k(z,Z)\,,
\end{align}
leading to the Lagrangian describing an infinite system:
\begin{align}
\mathcal{L}_{0,m}&=\sum_k(ia^*_k\dot{a}_k-\omega_ka^*_ka_k)\,.
\end{align}
In a finite wire of length $2L$, the general solution for a magnon of a given energy $\omega_k$ can be written as:
    \begin{align}
    \delta n(z,t)&=\sum_{k>0}a_k(t)\Psi_k(z,t)\,,\nonumber\\
\Psi_k(z,t)&=\alpha_k\psi_{k}(z,Z)+\beta_k\psi_{-k}(z,Z)\,,
\end{align}
with the coefficients $\alpha_k$ and $\beta_k$ determined from the assumption of vanishing spin current at the boundaries, ${\bs j}_s(\pm L,t)=0$. That is equivalent to imposing von Neumann boundary conditions, or 
\begin{align}
    \partial_z\delta n_{1,2}(z\rightarrow\pm L,t)=0\,,
\label{current_SM}    
\end{align} 
resulting in the following implicit equations:
\begin{align}
    &(\alpha_ke^{\pm ikL}+\beta_ke^{\mp ikL}){\rm sech}^2{(\pm L-Z)}+k\left[k(\alpha_ke^{\pm ikL}+\beta_ke^{\mp ikL})+i(\alpha_ke^{\pm ikL}-\beta_ke^{\mp ikL})\tanh{(\pm L-Z)}\right]=0\,.
\end{align}
When $|Z|\ll L$, we can ignore the term proportional to ${\rm sech}^2(\pm L-Z)$, and the boundary conditions can be simplified to:
\begin{align}
\alpha_k(k+i)e^{ikL}&+\beta_k(k-i)e^{-ikL} =0\,,\nonumber\\
\alpha_k(k-i)e^{-ikL}&+\beta_k(k+i) e^{ikL}=0\,,
\end{align}
from which we can extract the allowed wave-vectors $k_n\equiv k(n)$ and the associated eigen-energies $\omega_n=\omega(k_n)$. Specifically, we find the following transcendental equation
\begin{align}
\tan(2k_nL)=\frac{2k_n}{1-k_n^2}\,,    
\end{align}
which encodes two types of alternating solutions:  $\tan(k_nL)=k_n$ and $\tan(k_nL)=-1/k_n$, that yields  $\exp(2ik_nL)=\pm(i-k_n)/(i+k_n)$, respectively. Moreover, from
\begin{equation}
\beta_n=\alpha_n\frac{i+k_n}{i-k_n}e^{ik_nL}\equiv\pm\alpha_n
\end{equation}
we can  write
\begin{align}
    \Psi_n(z,Z)\approx C_n\left[\psi_{k_n}(z,Z)\pm \psi_{-k_n}(z,Z)\right]\equiv\pm\Psi_n^*(z,Z)\,, 
\end{align}
where $C_n\approx 1/\sqrt{2L}$ is found from the normalization of the wave-functions.
In the representation $\Psi_n(z,Z)$, the fluctuating magnonic fields can be formally written as: 
\begin{align}
\delta n_1(z,t)&=\frac{1}{2}\sum_n[\Psi_n(z,Z)a_{n}(t)+\Psi_n^*(z,Z)a^*_{n}(t)]\,,\nonumber\\
\delta n_2(z,t)&=\frac{i}{2}\sum_n[\Psi_n(z,Z)a_{n}(t)-\Psi_n^*(z,Z)a^*_{n}(t)]\,,
\end{align}
where $a_{n}(t)$ are the amplitudes of the spin waves with the quantum number $n$. 
Consequently, the free magnon Lagrangian becomes identical to that in Eq. (S27), with $k\rightarrow n$, with  the corresponding Hamiltonian
\begin{align}
H_{0,m}=\sum_{n} \omega_na_n^*a_n\,.
\end{align}

\begin{figure}[t]
\includegraphics[width=\linewidth]{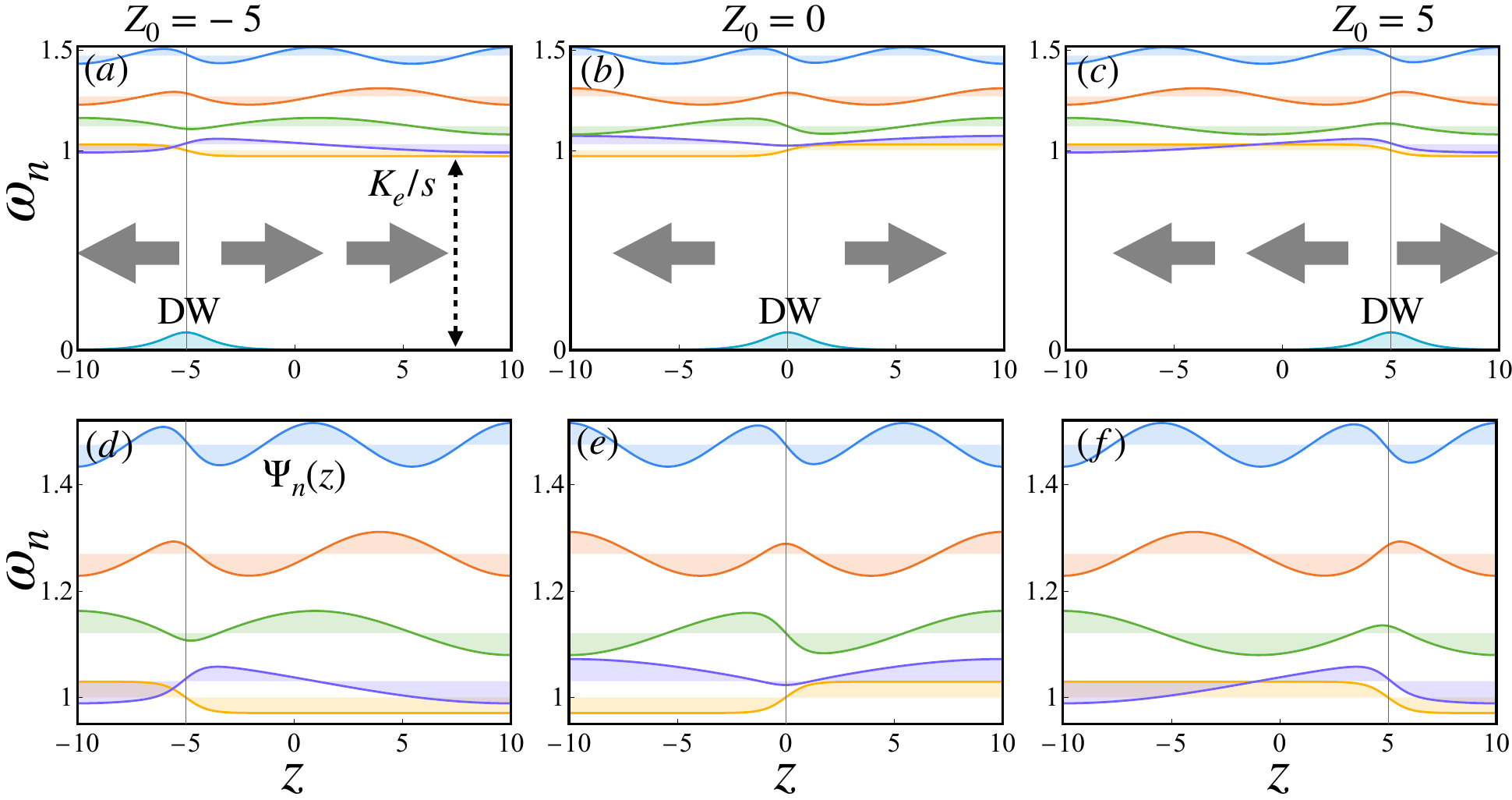}
\caption{The energies $\omega_n$ and the wave-functions $\Psi_n(z)$ of the first five magnonic modes ($n=1,\dots,5$) in the presence of a DW located at different positions $Z_0$ in the ferromagnetic wire of length $2L=20$. The arrows depict the spin orientation in each of the wire segment.  All energies are expressed in terms of the magnonic energy gap $\omega_0=K_e/s$.}
\label{fig2SM} 
\end{figure}

Let us briefly discuss the approximate spectrum at small momenta $k_n\ll1$ (that is, for wavelengths much larger than the DW size $\lambda$). In that case, we can write $\tan(2k_nL)\sim2k_n$ and, together with  $\tan(\theta+n\pi)=\tan(\theta)$, $n\in\mathcal{N}$, allows us to write $k_n=(r+n\pi/2)/L$, or 
\begin{equation}
\tan(2r)\sim 2r=2k\,.    
\end{equation}
Therefore, we find that the solutions are:
\begin{equation}
    k_n\approx\frac{\pi n}{2(L-1)}\,,
\end{equation}
with $n=0,1,2,\dots$. We can easily infer that $n=2k$ ($n=2k+1$), with $k=0,1,\dots$ pertaining to the first (second) type of magnon solution.

\subsection{Berry phase mediated interaction between the DW and magnons}

In the preceding section, the wave-functions $\Psi_n(z,Z)$ form a complete basis and can therefore be utilized to represent the magnonic fields $\delta n_{1,2}(z,t)$ also when $Z$ and $\Phi$ become time-dependent. 

The Berry phase contribution to the Lagrangian is
\begin{align}
    \mathcal{L}^B&=-s\int_{-L}^L dz\frac{{\bs d}\times{\bs n}}{1-{\bs n}\cdot{\bs d}}\,\dot{\bs n}\,,
\end{align}
where ${\bs d}$ is the (arbitrary) direction of the Dirac string. In the following, we choose ${\bs d}=-{\bs e}_z$, and expand $\mathcal{L}_B$ to the second order in the fluctuations. Retaining only the terms that contain the magnons, we obtain $\mathcal{L}_{\rm m}^B=\mathcal{L}^B_{{\rm m},1}+\mathcal{L}^B_{{\rm m},2}+\mathcal{L}^B_{{\rm int}}$ with
\begin{align}
\mathcal{L}^B_{{\rm m},1}&=s\int_{-L}^L dz\left(\delta n_1\dot{\Phi}\sin\theta_0+\delta n_2\dot{Z}\partial_z\theta_0\right)\,,\nonumber\\
    \mathcal{L}^B_{{\rm m},2}&=-s\int_{-L}^L dz\delta n_2\partial_t\delta n_1\,,\\
    \mathcal{L}^B_{{\rm int}}&=s\int_{-L}^L dz\left(-\dot{Z}\delta n_2\partial_Z\delta n_1+\frac{1}{2}(\delta n_1^2+\delta n_2^2)\cos\theta_0\dot{\Phi}\right)\equiv\dot{Z}P_{\rm m}^Z+\dot{\Phi}P_{\rm m}^\Phi\,,\nonumber
\end{align}
where
\begin{align}
    P_{\rm m}^Z&=-s\int_{-L}^L dz\delta n_2\partial_Z\delta n_1\,,\\
    P_{\rm m}^\Phi&=\frac{s}{2}\int_{-L}^L dz(\delta n_1^2+\delta n_2^2)\cos\theta_0\,.
\end{align}
Above, $P_{\rm m}^{Z}$ and $P_{\rm m}^{\Phi}$ represent the pressure of the magnons on the DW along the $Z$ and $\Phi$ directions, respectively \cite{KimPRB18}. These terms occur because the basis states become time-dependent: sliding and rotating along with the DW, respectively. The last term, in particular, pertain to a fictitious magnetic field $\dot{\Phi}$ along the wire axis, which in the case of a uniform magnet does not have any physical effects (it shifts the zero of energy by $\dot{\Phi}$). However, as demonstrated later, it can cause physical effects when the wire contains DW(s). Importantly, the linear coupling terms vanish due to the orthogonality of the magnon modes with the DW zero modes encoded by $\{Z,\Phi\}$.

Writing explicitly the expressions for the magnonic fields we find for the second term:
\begin{align}
    P_{\rm m}^\Phi&=\frac{s}{8}\sum_{n,m}\int_{-L}^Ldz([\Psi_n(z,Z)a_{n}(t)+\Psi_n^*(z,Z)a^*_{n}(t)][\Psi_m(z,Z)a_{m}(t)+\Psi_m^*(z,Z)a^*_{m}(t)]\nonumber\\
    &-[\Psi_n(z,Z)a_{n}(t)-\Psi_n^*(z,Z)a^*_{n}(t)][\Psi_m(z,Z)a_{m}(t)-\Psi_m^*(z,Z)a^*_{m}(t)])\cos(\theta_0)\approx\sum_{n,m}P_{nm}^\Phi a_n^*(t)a_m(t)\,,\\
    P_{nm}^\Phi&=\frac{s}{2}\int_{-L}^Ldz\Psi_n^*(z,Z)\Psi_m(z,Z)\cos(\theta_0)\,.
\end{align}
In the last line, we disregard the terms $\propto a_n(t)a_m(t)$ and $\propto a_n^*(t)a_m^*(t)$ as they relate to high-energy processes that are not effective in the low-energy theory describing the DW. Of particular importance in our work are the diagonal terms, that is 
\begin{align}
    P_{nn}^\Phi\equiv P_{n}(Z_0)&=\frac{s}{2}\int_{-L}^Ldz\Psi_n^*(z,Z)\Psi_n(z,Z)\cos(\theta_0)\,,
\end{align}
which will be discussed in more detail in the following sections.

\subsection{Equations of motion in the presence of magnons}

The equation of motion for the magnon field $a_n(t)$ reads (neglecting the Hamiltonian terms because they scale as $1/L$, as discussed later):
\begin{align}
i\dot{a}_n=\omega_na_n-\sum_m (\dot{Z}P^Z_{nm}+\dot{\Phi}P^\Phi_{nm})a_m\,,
\end{align}
while the DW dynamics in the absence of Gilbert damping becomes:
\begin{align}
\partial_\Phi U(\Phi)&-2s\dot{Z}+\dot{P}^\Phi_{\rm m}=0\,,\nonumber\\
2s\dot{\Phi}&+\frac{\partial V_{\rm pin}(Z)}{\partial Z}+\dot{P}_{\rm m}^Z=0\,.
\end{align}
Eliminating $\dot{Z}$ from the above equations:
\begin{equation}
    \dot{Z}=\frac{1}{2s}\left(\partial_\Phi U(\Phi)+\dot{P}_{\rm m}^\Phi\right)\,,
\end{equation}
we obtain
\begin{align}
M_\Phi\ddot{\Phi}&+\partial_\Phi U(\Phi)+\dot{P}_{\rm m}^\Phi
+\frac{M_{\Phi}}{2 SN_{dw}}\ddot{P}_{\rm m}^Z\approx M_\Phi\ddot{\Phi}+\partial_\Phi U(\Phi)+\dot{P}_{\rm m}^\Phi=0\,,
\end{align}
where in the last line we neglected the $P_{\rm m}^Z$ terms which become irrelevant for large wires and strongly pinned DW.
This in turn pertains to the total Lagrangian:
\begin{align}    
\mathcal{L}_{\rm tot}=\frac{M_{\Phi}\dot{\Phi}^2}{2}-U(\Phi)+\sum_{n}(i a_n^*\dot{a}_n-\omega_na_n^*a_n)+\dot{\Phi}P_{\rm m}^\Phi\,.
\end{align}
Dropping the $\Phi$ index from $P_{\rm m}^\Phi$ from here on and utilizing $P_\Phi=\partial_{\dot{\Phi}}\mathcal{L}_{\rm tot}=M_{\Phi}\dot{\Phi}+P_{\rm m}$, we can readily obtain the corresponding (quantum) Hamiltonian:
\begin{align}
    H_{\rm tot}=\frac{(P_\Phi-P_{\rm m})^2}{2M_{\Phi}}+\sum_{n}\omega_na_n^\dagger a_n+U(\Phi)\,,
\end{align}
which is the DW-magnon Hamiltonian displayed in the main text. 

\subsection{Uniform ferromagnetic lead coupled  to the magnetic cavity}

In a fixed frame, the magnetic lead is described by the Lagrangian
\begin{align}
    \mathcal{L}_{l}&=-\frac{1}{2}\int dz[2s\delta n_{l,1}\partial_t\delta n_{l,2}+\delta n_{l,1}\mathcal{H}_l\delta n_{l,1}+\delta n_{l,2}\mathcal{H}_l\delta n_{l,2}]\,,
\end{align}
where $n_{l,1(2)}$ are the amplitudes along the two orthogonal directions. More generally, the magnetization fluctuations read:
\begin{align}
    {\bs n}_l(z)={\bs e}_{1,l}n_{1,l}(z)+{\bs e}_{2,l}n_{l,2}(z)\,,
\end{align}
where
\begin{align}
    {\bs e}_{1,l}&=(-1,0,0);\,\,{\bs e}_{2,l}=(0,1,0)\,.
\end{align}
Consequently, the lead can be aligned either ferro- or anti-ferromagnetically with the cavity. The magnonic fields read (assuming Neumann boundary conditions at the interface with the cavity):
\begin{align}
    \delta n_{l}=\sum_{k\geq0}\sqrt{\frac{2}{L_l}}\cos{[k(z+L)]}b_k\,,
\end{align}
where $b_k$ are the amplitudes of the magnons with momentum $k\geq0$ and $L_l$ the size of the  lead.  The exchange coupling between the magnetic cavity and the lead at position $z=-L$ reads:
\begin{align}
    H_{\rm c}&=\mathcal{T}{\bs n}_l(-L)\cdot{\bs n}(-L)\approx\mathcal{T}({\bs e}_{1,l}\delta n_{1,l}+{\bs e}_{2,l}\delta n_{2,l})({\bs e}_{1}\delta n_{1}+{\bs e}_{2}\delta n_{2})=\frac{\mathcal{T}}{2}\left(\delta n_l^*\delta ne^{-i\Phi}+\delta n_l\delta n^*e^{i\Phi}\right)\,,
\end{align}
where $\mathcal{T}$ is the coupling strength, which we assumed to be isotropic, while $\delta n=\delta n_1+i\delta n_2$ (and similarly for the lead). Above, we kept only the leading-order contributions in the magnonic fields (the coupling along the $z$ direction only renormalizes the magnonic Hamiltonian for each ferromagnetic side with opposite signs for the two alignments). Performing the substitution:
\begin{align}
    \delta n\rightarrow\delta ne^{i\Phi}\,,
\end{align}
the tunneling Hamiltonian becomes independent on $\Phi$, but it affects the effective magnetic field felt by the magnons, $P_n=\dot{\Phi}P_{\rm 
 m}$, which now becomes:
\begin{align}
P_{\rm m}=\frac{s}{2}\int_{-L}^L dz(\cos{\theta}_0+1)(\delta n_1^2+\delta n_2^2)\,.
\end{align}
This leads to the modified total Hamiltonian, including the lead:
\begin{align}
    H_{\rm tot}=\frac{(P_\Phi-P_{\rm m})^2}{2M_{\Phi}}+\sum_{n}\omega_na_n^\dagger a_n+U(\Phi)+\sum_{k,n}\mathcal{T}_{kn}(b_k^\dagger a_n+{\rm h.c.}) +\sum_k\omega_kb_k^\dagger b_k\,,
\end{align}
and in the following, we retain only the Berry phase mediated interactions between the DW and the magnons (since the Hamiltonian terms scale as $1/L$ and are negligible for a large cavity). If we focus on the dynamics around one of the cavity modes, $\omega_n$, then $P_{\rm m}\approx P_{n}a_n^\dagger a_n$, with $P_{n}\equiv P_{n}(Z_0)$ being the result of integration for a given position $Z_0$  of the DW in the wire.

In Figs.~\ref{fig3SM}(a) and (b) we show $P_{n}$  as a function of $Z_0$ and $L$, respectively, for the first five magnonic modes. We see that the qualitative behavior follows:
\begin{align}
    P_n(Z_0)\approx1+Z_0/L\,,
\end{align}
up to corrections due to interference effects associated with the standing waves. However, this expression becomes exact for magnons that satisfy $k_nL\gg1$. 

\subsection{Hamiltonian induced DW-magnon couplings}

The Hamiltonian part of the Lagrangian can be found similarly, by expanding again up to second order in the fluctuations $\delta n_{1,2}$. Retaining only the second order corrections in the magnonic fields (since the first order correspond to high-energy processes), the interaction with the DW becomes: 
\begin{align}
\mathcal{L}^H_{int}&=\frac{1}{2}\int_{-L}^L dz\left(\delta n_1V_1\delta n_1+\delta n_2V_2\delta n_2+\delta n_1V_3\delta n_2\right)\,,\nonumber\\
V_1&=K_h\cos(2\theta_0)\sin^2\Phi-\frac{b}{2}\sin\theta_0\sin\Phi\,,\\
V_2&=K_h(\cos^2\Phi-\sin^2\theta_0\sin^2\Phi)-\frac{b}{2}\sin\theta_0\sin\Phi\,,\nonumber\\
V_3&=K_h\cos\theta_0\sin(2\Phi)\nonumber\,.   
\end{align}
We can further write:
\begin{align}
    \mathcal{L}^H_{int}&\approx\frac{1}{4}\sum_{n,m}\int_{-L}^L dz\Psi_n^*(z,Z)\Psi_m(z,Z)(V_1+V_2)a_n^*a_m\nonumber\\
    &\equiv\sum_{n,m}(K_hJ_{nm}\sin^2\Phi+bF_{nm}\sin\Phi)a_n^*a_m\,,\\
    J_{nm}\equiv J_{nm}(Z_0)&=-\frac{3}{4}\int_{-L}^Ldz\Psi_n^*(z,Z)\Psi_m(z,Z)\sin^2\theta_0\,,\nonumber\\
    F_{nm}\equiv F_{nm}(Z_0)&=\frac{1}{8}\int _{-L}^Ldz\Psi_n^*(z,Z)\Psi_m(z,Z)\sin\theta_0\,,\nonumber
\end{align}
where we disregarded terms  containing $a_n^*a_m^*$ and $a_na_m$ as they do not conserve the number of magnons and correspond to high energy process (note that $V_3$ only contains such types of process, and therefore does not contribute to the low-energy theory). Furthermore, we also disregarded the terms that only affect the magnons independently on the DW degrees of freedom. 

\begin{figure}[t]
\includegraphics[width=0.8\linewidth]{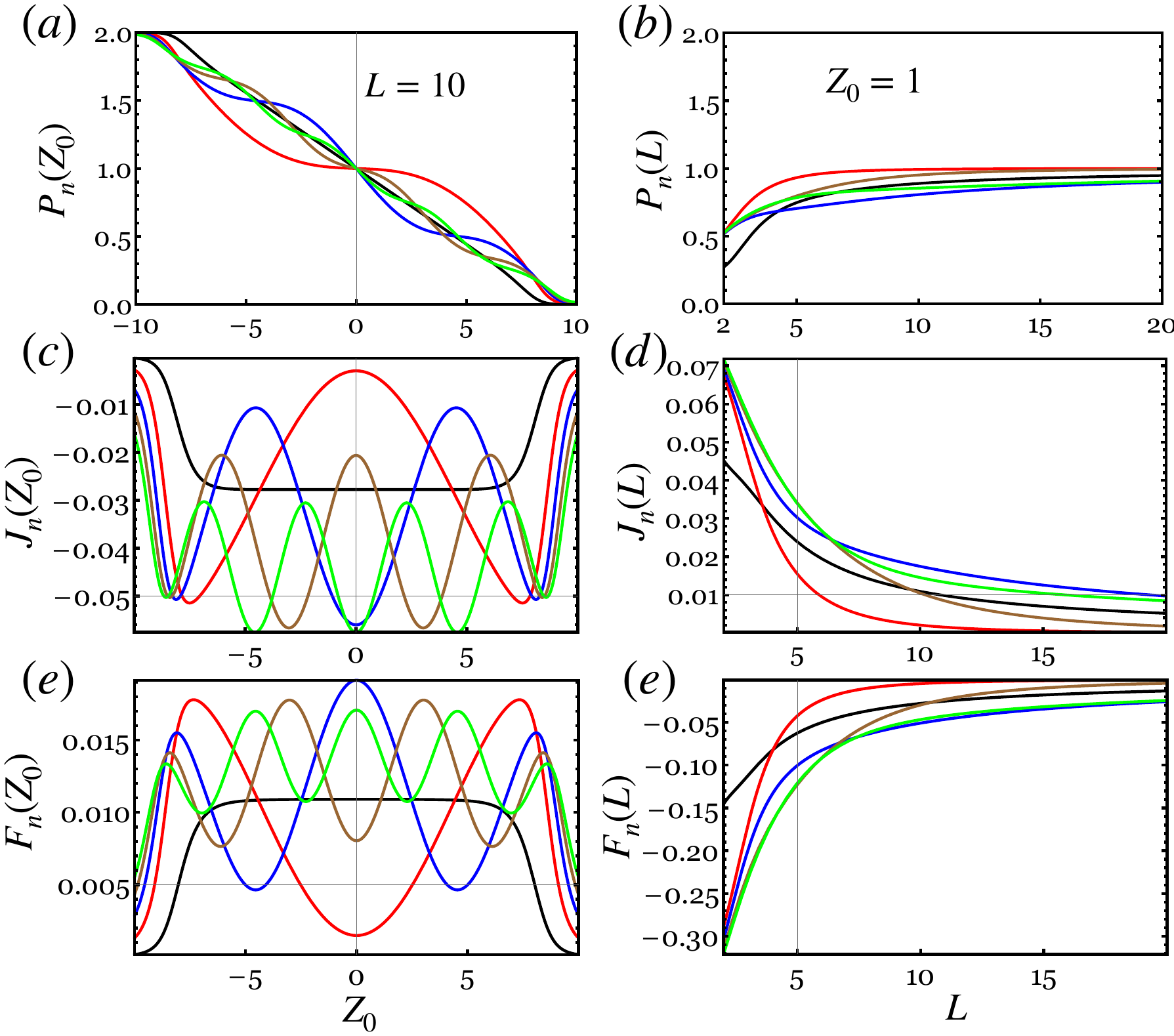}
\caption{(a) The Berry phase accumulated by the first five magnon modes for $L=10$ as a function of the DW position $Z_0$ (b) Same, as a function of $L$ for $Z_0=1$. The black, red, blue, brown, and green curves correspond, in the increasing order, to the $n=1$ (uniform mode) to $n=5$ magnonic modes in the wire. They all exhibit the same behavior: the Berry phase is maximum (minimum) at $Z_0=-L$ ($Z_0=L$), while away from $Z_0=0$ they show oscillations due to interference effects. (c)--(d) The Hamiltonian terms $J_n$ and $F_n$ as a function of $Z_0$ and $L$, respectively.}
\label{fig3SM} 
\end{figure}

In Figs.~\ref{fig3SM}(c)-(f) we show the dependence of the diagonal terms $J_{nn}\equiv J_n(L,Z_0)$ and $F_{nn}\equiv F_n(L,Z_0)$ on both the position of the DW ($Z_0$) and the size of the wire ($L$). We see that both $J_n(L, Z_0),F_n(L, Z_0)\propto1/L$, and thus become negligible for $L\gg1$ compared to the Berry phase contribution.  

\subsection{Effect of the pinning potential on the magnons}

Here we scrutinize the effect of the pinning potential on the magnons. Assuming that the DW is strongly pinned at $Z_0$, the scattering of the magnons can be extracted by solving the pinning potential at position $Z_0$. That is, we need to solve the following Schrodinger equation: 
\begin{align}
    [\mathcal{H}_0+V_0\delta(z-Z_0)]\Psi_n(z,Z_0)=\omega_n\Psi_n(z,Z_0)\,,
\end{align}
where the (dimensionless) scattering strength is $V_0=\delta K_e/(2K_e)$, and the wave-function is  subject to the boundary conditions $\partial_z{\Psi}_{n}(z=\pm L)=0$ (vanishing spin current). We seek for solutions of the form:
\begin{align}
    \Psi_{n}(z)=\left\{
    \begin{array}{cc}
    \alpha_{L,k}\psi_k(z,Z_0)+\beta_{L,k}\psi_{-k}(z,Z_0)& -L<z<Z_0\,,\\\\
    \alpha_{R,k}\psi_k(z,Z_0)+\beta_{R,k}\psi_{-k}(z,Z_0)&Z_0<z<L\,,
    \end{array}
    \right.
\end{align}
so that the connection between the two across the scatterer reads:
\begin{align}
    -[\Psi'_n(Z_0^+,Z_0)-\Psi'_n(Z_0^-,Z_0)]+V_0\Psi_{n}(Z_0,Z_0)=0\,,
\end{align}
as well as $\Psi_n(Z_0^+,Z_0)=\Psi_n(Z_0^-,Z_0)$, stemming from the continuity of the wave-function. Hence, the equations determining the spectrum and the wave-functions, respectively, read: 
\begin{align}
\alpha_{k,R}(k+i)e^{ikL}&+\beta_{k,R}(k-i)e^{-ikL} =0\,,\nonumber\\
\alpha_{k,L}(k-i)e^{-ikL}&+\beta_{k,L}(k+i)e^{ikL}=0\,,\\
(\alpha_{k,L}&-\alpha_{k,R})e^{ikZ_0}-(\beta_{k,L}-\beta_{k,R})e^{-ikZ_0}=0\,,\nonumber\\
(\alpha_{k,L}&-\alpha_{k,R})e^{ikZ_0}+(\beta_{k,L}-\beta_{k,R})e^{-ikZ_0}+\frac{ikV_0}{1+k^2}(\alpha_{L,k}e^{ikZ_0}-\beta_{L,k}e^{-ikZ_0})=0\nonumber\,.
\end{align}
The resulting eigen-momentum equation becomes:
\begin{align}
    2(1+k^2)\left[2k\cos(2kL)-(1-k^2)\sin(2kL)\right]-kV_0\left[(1-k^2)\cos(2kL)+2k\sin(2kL)-(1+k^2)\cos(2k Z_0)\right]=0\,.
\end{align}
Although we cannot find an exact solution for $k_n$ analytically, we can determine approximate solutions in the limit $L\rightarrow\infty$. We can write
\begin{align}
    k_n\approx\frac{1}{L}\left(\frac{n\pi}{2}+a_n\right)\,,
\end{align}
with $n=0,1,\dots$, and  $a_n$ needs to be found perturbatively. In doing so, we find:
\begin{align}
    a_n=\frac{n\pi}{8L}[4-V_0(1+(-1)^n\cos(\pi n Z_0/L)]\,.
\end{align}
Hence, we can deduce that magnons experience minimal scattering at low $k_n$ values, and the impact of potential pinning can be safely ignored.

\section{Derivation of the reduced density matrix describing the DW evolution}

In this section, we provide additional information regarding the dynamics of the domain wall (DW) in the presence of driven magnons in the finite magnetic wire system.

\subsection{Linearised wire Hamiltonian in the original frame}

We assume that the magnetic cavity is driven at a frequency $\Omega_n\sim\omega_n$. In a frame rotating at this frequency, the total Hamiltonian reads: 
\begin{align}
    H_{\rm tot}=\frac{(P_\Phi-P_{n}a_n^\dagger a_n)^2}{2M_{\Phi}}+U(\Phi)+\Delta_n a^\dagger_n a_n+h_{n}( a_n^\dagger+a_n)\,,
\end{align}
where $\Delta_n=\omega_n-\Omega_n$ and  $h_{n}=\sqrt{W_n\kappa_{n,ex}/\omega_n}$ is the  amplitude of the $ac$ magnetic field driving the mode $n$ from the left port, with $\kappa_{n,ex}$ being the corresponding decay rate and $W_n$ the injected power. Note that here we disregarded the other (non-driven) magnonic modes, as they are not resonant with the DW states and therefore do not participate in the dynamics. The evolution of the combined system density matrix $\rho_{\rm S}$ can be captured by the following Lindblad equation:
\begin{align}
    \dot{\rho}_{\rm S}=-i[H_{\rm tot},\rho_{\rm S}]&+\kappa_n\mathcal{D}(a_n)\rho_{\rm S}+\sum_{j,k}\gamma_{jk}\mathcal{D}(A_{jk})\rho_{\rm S}\,,
\end{align}
where we recall that $A_{jk}=|j\rangle\langle k|$. Next, we can expand the magnonic cavity field operators around their steady-state value, $\alpha_n$, that is, $a_n\rightarrow\alpha_n+a_n$. The average value $\alpha_n$ is found by requiring that the linear terms in $a_n$ and $a_n^\dagger$ that do not involve DW operators vanish, leading to the following equation
\begin{align}    
h_n=\alpha_n\left[i\kappa_n-\left(\Delta_n+\frac{P^2_{n}|\alpha_n|^2}{M_\Phi}\right)\right]\,.
\end{align}
The resulting density matrix equation is the same as above, but the unitary part is described by the Hamiltonian 
\begin{align}
    H_{\rm tot}&=\frac{(P_\Phi-P_n|\alpha_n|^2)^2}{2M_\Phi}-\frac{P_nP_\Phi(\alpha_na_n^\dagger+\alpha_n^*a_n)}{M_\Phi}+\underbrace{\left(\Delta_n-\frac{P^2_{n}|\alpha_n|^2}{M_\Phi}\right)}_{\Delta_n'}a_n^\dagger a_n+U(\Phi)\,,
\end{align}
In the limit of small power, or $|\alpha_n|\ll1$, we can disregard all terms $\propto|\alpha_n|^2$ above, which results in the Hamiltonian in Eq.~(3) in the main text.

\subsection{Total wire Hamiltonian in the rotated  frame}

The gauge field introduced by the magnons can be removed by a unitary transformation $\mathcal{U}(\Phi)=e^{iP_n\Phi}$, so that the total Hamiltonian becomes:
\begin{align}
    \widetilde{H}_{\rm tot}\equiv \mathcal{U}^\dagger(\Phi)H_{\rm tot}\mathcal{U}(\Phi)=\frac{P_\Phi^2}{2M_{\Phi}}+U(\Phi)+\Delta_n a^\dagger_n a_n+h_{n}(a_n^\dagger e^{-iP_n\Phi}+a_ne^{iP_n\Phi})\,,
\end{align}
which shows that the entire effect of the coupling is lumped into the driving term. Note that in effect this transformation also alters the DW dissipator. However, since the rates are already small compared to the level spacing, we disregard any such modifications induced by this transformation. 
We can further displace the magnons by a quantity that, in turn, removes the linear terms that are independent of $\Phi$. That is:
\begin{align}
    \widetilde{H}_{\rm tot}&=\frac{P_\Phi^2}{2M_{\Phi}}+U'(\Phi)+\Delta_n a^\dagger_n a_n+h_{n}\left[a_n^\dagger(e^{-iP_n\Phi}-1)+a_n(e^{iP_n\Phi}-1)\right]\,,\nonumber\\
U'(\Phi)&=U(\Phi)+h_{n}\left[\alpha_n^*(e^{-iP_n\Phi}-1)+\alpha_n(e^{iP_n\Phi}-1)\right]\,.
\end{align}
In the limit $\kappa_n\ll\Delta_n$ (which is needed for efficient cooling), we can approximate $h_n\approx-\alpha_n\Delta_n$, so that:
\begin{align}
    \widetilde{H}_{\rm tot}&=\frac{P_\Phi^2}{2M_{\Phi}}+U'(\Phi)+\Delta_n a^\dagger_n a_n-\alpha_n\Delta_n\left[a_n^\dagger(e^{-iP_n\Phi}-1)+a_n(e^{iP_n\Phi}-1)\right]\,,\nonumber\\
U'(\Phi)&=U(\Phi)-2\bar{n}\Delta_n\left[\cos(P_n\Phi)-1\right]\,,
\end{align}
where $\bar{n}=|\alpha_n|^2$ is the average number of magnons in mode $\omega_n$. As in the previous subsection, if $|\alpha_n|\ll1$, we can disregard the modifications of the potential $U(\Phi)$ by the magnons.

\subsection{Effective DW density matrix evolution}

The total density matrix evolution of the combined  system in the lab frame is described by the following  Liouville equation:
\begin{align}
    \dot{\rho}_{\rm S}&=(\mathcal{L}_{\rm m}+\mathcal{L}_{\Phi}+\mathcal{L}_{\Phi-{\rm m}})\rho_{\rm S}\,,
\end{align}
with
\begin{align}
    \mathcal{L}_{\rm m}\rho_{\rm S}&=-i\Delta_{n}[a_n^\dagger a_n,\rho_{\rm S}]+\kappa_n\mathcal{D}(a_n)\rho_{\rm S}\,,\nonumber\\
    \mathcal{L}_{\Phi}\rho_{\rm S}&=-i[H_{\Phi},\rho_{\rm S}]+\sum_{j,k}\gamma_{jk}\mathcal{D}(A_{jk})\rho_{\rm S}\,,\\   
    \mathcal{L}_{\Phi-{\rm m}}\rho_{\rm S}&=-ig_n[P_\Phi  X_n,\rho_{\rm S}]\,,\nonumber
\end{align}
where $X_n=(\alpha_na_n^\dagger+\alpha_n^*a_n)/|\alpha_n|$ and we recall that  $g_n=|\alpha_n| P_n(Z_0)/M_\Phi$. 
Let us define $\rho_\Phi(t)={\rm Tr}_{\rm m}[\rho_{\rm S}(t)]$ as the reduced density matrix that describes the dynamics of the DW only, and adiabatically eliminate the magnons. Then, in the interaction picture and within the Born-Markov approximation we obtain \cite{gardiner00}:
\begin{align}
    \dot{\rho}_\Phi(t)&=
    -g_n^2\int_0^\infty d\tau{\rm Tr}_{\rm m}\{[P_\Phi(t)X_n(t),[P_\Phi(t-\tau)X_n(t-\tau),\rho_\Phi(t)\otimes\rho_{\rm m}]]\}\nonumber\\
    &=-g_n^2\left(\int_0^\infty d\tau\langle X_n(t)X_n(t-\tau)\rangle[P_\Phi(t),[P_\Phi(t-\tau),\rho_\Phi(t)]]+\int_0^\infty d\tau\langle[X_n(t),X_n(t-\tau)]\rangle[P_\Phi(t),\rho_\Phi(t) P_\Phi(t-\tau)]\right)
\end{align}
where $P_\Phi(t)=\sum_{j,k}\langle j|P_\Phi|k\rangle e^{i(\epsilon_j-\epsilon_k)t}A_{jk}$, and \cite{gardiner00} 
\begin{align}
\langle X_n(t)X_n(t-\tau)\rangle&={\rm Tr}_{\rm m}[X_ne^{\mathcal{L}_{\rm m}\tau}X_n\rho_{\rm m}(t)]\equiv\langle a_n^\dagger(t)a_n(t-\tau)\rangle+\langle a_n(t)a_n^\dagger(t-\tau)\rangle\,,\nonumber\\
\langle [X_n(t),X_n(t-\tau)]\rangle&={\rm Tr}_{\rm m}[X_ne^{\mathcal{L}_{\rm m}\tau}[X_n,\rho_{\rm m}(t)]]\equiv\langle [a_n^\dagger(t),a_n(t-\tau)]\rangle+\langle[a_n(t),a_n^\dagger(t-\tau)]\,,
\end{align}
with $\rho_{\rm m}$ being the (thermal) density matrix describing the magnons in mode $\omega_n$ around the steady state solution $\alpha_n$. The correlators above can be evaluated by employing the quantum regression theorem \cite{gardiner00}, which gives:
\begin{align}
  \langle a_n^\dagger(t)a_n(t-\tau)\rangle&=e^{-(-i\Delta_n+\kappa_n/2)\tau}\langle a_n^\dagger(t)a_n(t)\rangle=e^{-(-i\Delta_n+\kappa_n/2)\tau}n_B(\omega_n)\,,\nonumber\\
    \langle a_n(t)a_n^\dagger(t-\tau)\rangle&=e^{-(i\Delta_n+\kappa_n/2)\tau}\langle a_n(t)a_n^\dagger(t)\rangle=e^{-(i\Delta_n+\kappa_n/2)\tau}[1+n_B(\omega_n)]\,,\\
    \langle a_n(t-\tau)a_n^\dagger(t)\rangle&=e^{-(-i\Delta_n+\kappa_n/2)\tau}\langle a_n(t)a_n^\dagger(t)\rangle=e^{-(-i\Delta_n+\kappa_n/2)\tau}[1+n_B(\omega_n)]\,,\nonumber\\
    \langle a_n^\dagger(t-\tau)a_n(t)\rangle&=e^{-(i\Delta_n+\kappa_n/2)\tau}\langle a_n^\dagger(t)a_n(t)\rangle=e^{-(i\Delta_n+\kappa_n/2)\tau}n_B(\omega_n)\nonumber\,.
\end{align}
Therefore:
\begin{align}
    \langle X_n(t)X_n(t-\tau)\rangle&=\left[e^{-i\Delta_n\tau}(1+n_B(\omega_n))+e^{i\Delta_n\tau}n_B(\omega_n)\right]e^{-\kappa_n\tau/2}\,,\nonumber\\
    \langle [X_n(t),X_n(t-\tau)]\rangle&=\left(e^{i\Delta_n\tau}-e^{-i\Delta_n\tau}\right)e^{-\kappa_n\tau/2}\,.
\end{align}
We can insert the above equations in the expression for the density matrix evolution to obtain
\begin{align}
\dot{\rho}_\Phi(t)=&-g_n^2\sum_{j,k}|\langle j|P_\Phi|k\rangle|^2\left(\int_0^\infty d\tau e^{-i(\epsilon_j-\epsilon_k)\tau}\left[e^{-i\Delta_n\tau}(1+n_B(\omega_n))+e^{i\Delta_n\tau}n_B(\omega_n)\right]e^{-\kappa_n\tau/2}[A_{jk},[A_{jk}^\dagger,\rho_\Phi(t)]]\right.\nonumber\\
    &\left.+\int_0^\infty d\tau e^{-i(\epsilon_j-\epsilon_k)\tau}\left(e^{-i\Delta_n\tau}-e^{i\Delta_n\tau}\right)e^{-\kappa_n\tau/2}[A_{jk},\rho_\Phi(t) A_{jk}^\dagger]\right)\nonumber\\
    &=g_n^2\sum_{j,k}|\langle j|P_\Phi|k\rangle|^2\bigg(\frac{\kappa_n}{2}\left[\frac{1+n_B(\omega_n)}{(\epsilon_k-\epsilon_j-\Delta_n)^2+(\kappa_n/2)^2}+\frac{n_B(\omega_n)}{(\epsilon_k-\epsilon_j+\Delta_n)^2+(\kappa_n/2)^2}\right]\mathcal{D}[A_{jk}]\rho_\Phi(t)\nonumber\\
    &+i\left[\frac{(1+n_B(\omega_n))(\epsilon_k-\epsilon_j-\Delta_n)}{(\epsilon_k-\epsilon_j-\Delta_n)^2+(\kappa_n/2)^2}+\frac{n_B(\omega_n)(\epsilon_k-\epsilon_j+\Delta_n)}{(\epsilon_k-\epsilon_j+\Delta_n)^2+(\kappa_n/2)^2}\right][A_{jk}^\dagger A_{jk},\rho_\Phi(t)]\bigg)\nonumber\\
    &\equiv i\sum_j(\epsilon_j+\delta\epsilon_j)[A_{jk}^\dagger A_{jk},\rho_\Phi(t)]+\sum_{j,k}(\Gamma_{jk}^n+\gamma_{jk})\mathcal{D}[A_{jk}]\rho_\Phi(t)\,,
\end{align}
where 
\begin{align}
\Gamma^n_{jk}&=\frac{g_n^2\langle j|P_\Phi|k\rangle|^2\kappa_n}{2}\left[\frac{1+n_B(\omega_n)}{(\epsilon_k-\epsilon_j-\Delta_n)^2+(\kappa_n/2)^2}+\frac{n_B(\omega_n)}{(\epsilon_k-\epsilon_j+\Delta_n)^2+(\kappa_n/2)^2}\right]\approx\frac{1}{2}\frac{g_n^2|\langle j|P_\Phi|k\rangle|^2\kappa_n}{(\epsilon_k-\epsilon_j-\Delta_n)^2+(\kappa_n/2)^2}\,,\\ \delta\epsilon^n_{j}&=g_n^2\sum_k|\langle j|P_\Phi|k\rangle|^2\left[\frac{(1+n_B(\omega_n))(\Delta_n-\epsilon_k+\epsilon_j)}{(\epsilon_k-\epsilon_j-\Delta_n)^2+(\kappa_n/2)^2}+\frac{n_B(\omega_n)(\Delta_n+\epsilon_k-\epsilon_j)}{(\epsilon_j-\epsilon_k+\Delta_n)^2+(\kappa_n/2)^2}\right]\nonumber\\
&\approx\sum_k\frac{g_n^2|\langle j|P_\Phi|k\rangle|^2(\Delta_n-\epsilon_k+\epsilon_j)}{(\epsilon_k-\epsilon_j-\Delta_n)^2+(\kappa_n/2)^2}\,.
\end{align}
In the last lines of each above equation, we assumed $k_BT\ll\omega_n$ such that $n_B(\omega_n)\approx0$. 
Note that $\delta\epsilon_j$ is the Lamb shift of the DW levels due to interaction with the magnons. We can now establish the stationary density matrix pertaining to the DW, that is, $\dot{\rho}_\Phi=0$, and $\rho_\Phi=\sum_{j}p_j|j\rangle\langle j|$, with $p_j$ the (out-of-equilibrium) occupation probability of the DW state of energy $\epsilon_j$. They obey the following detailed balance equation:
\begin{align}
    p_j\sum_{k\neq j}\underbrace{(\gamma_{kj}+\Gamma^n_{kj})}_{\displaystyle{\Gamma_{kj}^{\rm tot}}}-\sum_{k\neq j}p_k\underbrace{(\gamma_{jk}+\Gamma^n_{jk})}_{\displaystyle{\Gamma_{jk}^{\rm tot}}}=0\,,
\end{align}
and subject to the normalization condition $\sum_{j}p_j=1$. In the main text, we find $p_j$ numerically by including the first 20 DW levels in the above equation. 

\subsection{Wigner function (or quasi-probability distribution)}

The Wigner function is defined as \cite{Harochebook}:
\begin{align}
    W(\Phi,P_\Phi)=\frac{1}{\pi}\int d\Phi'\langle\Phi-\Phi'|\rho_\Phi|\Phi+\Phi'\rangle e^{2iP_{\Phi}\Phi'}\,,
\end{align}
and it contains all possible information
about the DW state. The density operator $\rho_\Phi$ can be deduced from it and hence the expectation value of all the possible DW observables. 
Its physical interpretation can be traced back to the expectation values of the operators. Let $A $ be an operator in the phase space $\Phi-P_\Phi$. Then, its expectation value reads:
\begin{align}
    \langle A\rangle=\int\int d\Phi d P_\Phi W(\Phi,P_\Phi)A\,.
\end{align}
Although naively one could interpret $W(\Phi,P_\Phi)$ as a probability distribution in the phase space, this is not, in fact, a classical probability distribution. A feature that distinguishes the Wigner function from a classical probability distribution is the fact that the former can take negative values. Generally, negativities in the Wigner function are a good indication of the nonclassical nature of a quantum state \cite{Harochebook}. 

At the origin, i.e. $\Phi=P_\Phi=0$, we find:
\begin{equation}
    W(0,0)=\frac{1}{\pi}\int d\Phi'\langle-\Phi'|\rho_\Phi|\Phi'\rangle=\frac{1}{\pi}\sum_jp_j\int d\Phi'\Psi_j^{*}(-\Phi')\Psi_j(\Phi')\equiv\frac{1}{\pi}\sum_{j}(-1)^{j+1}p_j\,,
\end{equation}
which is the expression employed in the main text. 

\subsection{Detection of the DW levels occupation from the emitted power spectrum}

In experimental settings, the population of the DW levels can be determined by examining the power spectrum of the magnons that are emitted. This involves applying a separate laser, which is weakly driven, to the magnetic cavity at one of the ports. This laser is tuned to resonate with one of the magnonic modes, denoted as $\omega_p$ (distinct from the laser used to manipulate occupations). The resulting magnonic radiation is subsequently analyzed.

Let us establish the equations of motion for the quantum fluctuations of the cavity operators $a_p$s around the average value $\alpha_p$:
\begin{align}
    \dot{a}_p=-\frac{\kappa_p}{2}a_p&-i\frac{P_n(Z_0)}{M_\Phi} P_\Phi(a_p+\alpha_p)+\sqrt{\kappa_{p,ex}}a_{p,\rm in}(t)+\sqrt{(\kappa_{p}-\kappa_{p,ex})}b_{p,\rm in}(t)\,,
\end{align}
where $\kappa_p$ and $\kappa_{p,ex}$ are the total decay rate and the decay rate associated with the vacuum noise of the laser field, respectively. Note that we can write the input contribution as $a_{p,\rm in}(t)\rightarrow h_p(t)+a_{p,\rm in}$, which is the sum of the coherent part of the input and fluctuations, respectively. The output power spectrum associated with a mode driven by the laser frequency $\omega_L=\omega_p$ is given by \cite{WilsonRae_NJP_08}:
\begin{align}
    \mathcal{S}_p(\omega)=\frac{1}{2\pi}\int_{-\infty}^\infty d\tau e^{-i(\omega-\omega_L)\tau}\langle a_{p, \rm out}^\dagger(t+\tau)a_{p,\rm out}(t)\rangle\,, 
\end{align}
where the output field, $a_{p,\rm out}(t)$, is related to the input one, $a_{p,\rm in}(t)$, via the input-output relations:
\begin{align}
   a_{p,\rm out}(t)=\sqrt{\kappa_{p,ex}}[a_p(t)+\alpha_p]+h_{p}(t)+a_{p,\rm in}(t)\,, 
\end{align}
and $\langle\dots\rangle$ represents the expectation value over the stationary state. We are hence left with finding the cavity field evolution, which can be formally integrated as:
\begin{align}
    a_p(t)&=a_{p,0}(t)+i\frac{P_n(Z_0)}{M_\Phi}\int_0^t d\tau e^{-\kappa_p(t-\tau)/2}P_\Phi(\tau)[a_p(\tau)+\alpha_p]\nonumber\\
    &\approx a_{p,0}(t)+i\frac{P_n(Z_0)}{M_\Phi}\sum_{j,k}\langle j|P_\Phi|k\rangle\int_0^t d\tau e^{-\kappa_p(t-\tau)/2}e^{-i(\epsilon_j-\epsilon_k)(t-\tau)}A_{jk}(t)[a_p(\tau)+\alpha_p]\nonumber\\
    &\approx a_{p,0}(t)-\underbrace{\frac{\alpha_pP_n(Z_0)}{M_\Phi}}_{\displaystyle g_p}\sum_{j,k}\frac{\langle j|P_\Phi|k\rangle}{\epsilon_j-\epsilon_k-i\kappa_p/2}A_{jk}(t)\,,
\end{align}
where $a_{p,0}(t)$ is the solution in the absence of the coupling to the DW, i.e. for $P_n(Z_0)=0$, and we kept the leading order terms in the Dyson series.  Hence, the noise can be written as:
\begin{align}
    \mathcal{S}_p(\omega)&=\frac{1}{2\pi}\int_{-\infty}^\infty d\tau e^{-i(\omega-\omega_L)\tau}\bigg\langle\left[ \sqrt{\kappa_{p,ex}}a_p^\dagger(t+\tau)+a_{p,\rm in}^\dagger(t+\tau)+\sqrt{\kappa_{p,ex}}\alpha_p^*-\frac{ih_p^*}{\sqrt{k_{p,ex}}}\right]\nonumber\\
    &\times\left[\sqrt{\kappa_{p,ex}}a_p(t)+a_{p,\rm in}(t)+\sqrt{\kappa_{p,ex}}\alpha_p+\frac{ih_p}{\sqrt{\kappa_{p,ex}}}\right]\bigg\rangle\,.
\end{align}
This can be expressed as the sum of elastic scattering (which does not involve DW) and the inelastic contribution (or side-band transitions), that is,  $S_p(\omega) \equiv S_{p,\rm el}(\omega)+S_{p,\rm in}(\omega)$. The former is given by:
\begin{align}
    &\mathcal{S}_{p,\rm el}(\omega)\approx\frac{1}{2\pi}\int_{-\infty}^\infty d\tau e^{-i(\omega-\omega_L)\tau}\left[\frac{1}{\kappa_{p,ex}}|h_p|^2+\kappa_{p,ex}|\alpha_p|^2+i(\alpha_p^*h_p-\alpha_ph_p^*)\right]\nonumber\\
    &=\frac{|h_p|^2}{\pi\kappa_{p,ex}}\int_{-\infty}^\infty d\tau e^{-i(\omega-\omega_L)\tau}\left[1+\frac{8\kappa_{p,ex}^2}{\kappa_p^2}-\frac{8\kappa_p\kappa_{p,ex}}{\kappa_p^2}\right]=\frac{W_p}{\omega_p}\left[1-\frac{8\kappa_{p,ex}(\kappa_p-\kappa_{p,ex})}{\kappa_p^2}\right]\delta(\omega-\omega_L)\,,
\end{align}
with the delta function highlighting the elastic processes. The second term, which takes on the behavior of the DW, is expressed as follows:
\begin{align}    
\mathcal{S}_{p,\rm in}(\omega)&=\frac{\kappa_{p,ex}}{2\pi}\int_{-\infty}^\infty d\tau e^{-i(\omega-\omega_L)\tau}\langle a_p^\dagger(t+\tau)a_p(t)\rangle=\frac{\kappa_{p,ex}}{2\pi}{\mathcal Re}\int_{0}^\infty d\tau e^{-i(\omega-\omega_L)\tau}\langle a_p^\dagger(t+\tau)a_p(t)\rangle\nonumber\\
&=-\frac{\kappa_{p,ex}}{2\pi}g_p^2{\mathcal Re}\int_{0}^\infty d\tau e^{-i(\omega-\omega_L)\tau}\bigg\langle\int_0^{t+\tau}d\tau_1e^{-\kappa_p(t+\tau-\tau_1)/2}P_\Phi(\tau_1)\int_0^{t}d\tau_2e^{-\kappa_p(t-\tau_2)/2}P_\Phi(\tau_2)\bigg\rangle\nonumber\\
&\approx\frac{\kappa_{p,ex}}{2\pi}g_p^2{\mathcal Re}\sum_{j,j';k,k'}|\langle j|P_\Phi|k\rangle|^2\int_{0}^\infty d\tau e^{-i(\omega-\omega_L)\tau}
\bigg\langle\int_0^\infty d\tau_1e^{-\kappa_p\tau_1/2}e^{-i(\epsilon_j-\epsilon_k)\tau_1}A_{jk}(t+\tau)\nonumber\\
&\times\int_0^\infty d\tau_2e^{-\kappa_p\tau_2/2}e^{-i(\epsilon_{j'}-\epsilon_{k'})\tau_2}A_{j'k'}(t)\bigg\rangle\nonumber\\
&=\frac{\kappa_{p,ex}}{2\pi}g_p^2{\mathcal Re}\int_{0}^\infty d\tau e^{-i(\omega-\omega_L)\tau}\sum_{j,k}\frac{|\langle j|P_\Phi|k\rangle|^2}{(\epsilon_j-\epsilon_k)^2+(\kappa_p/2)^2}\langle A_{jk}(t+\tau)A_{kj}(t)\rangle\nonumber\\
&=\frac{\kappa_{p,ex}}{2\pi}g_p^2{\mathcal Re}\int_{0}^\infty d\tau e^{-i(\omega-\omega_L)\tau}\sum_{j,k}\frac{|\langle j|P_\Phi|k\rangle|^2}{(\epsilon_j-\epsilon_k)^2+(\kappa_p/2)^2}e^{-M_{jk}\tau}p_j\nonumber\\
&=\frac{\kappa_{p,ex}}{2\pi}g_p^2\sum_{j,k}\frac{|\langle j|P_\Phi|k\rangle|^2}{(\epsilon_j-\epsilon_k)^2+(\kappa_p/2)^2}\frac{\Gamma_{jk}^{\rm eff}}{(\omega-\omega_L-\epsilon_j+\epsilon_k)^2+(\Gamma_{jk}^{\rm eff})^2}p_j\,,
\end{align}
where $\Gamma_{jk}^{\rm eff}={\mathcal  Im}M_{jk}=\sum_p(\Gamma_{pk}^n+\Gamma_{pj}^n+\gamma_{pk}+\gamma_{pj})=\Gamma_{kj}^{\rm eff}$, as demonstrated below. Therefore, for $\omega\neq\omega_L$, the total contribution to the noise is due to side-band transitions, as mentioned in the main text.  

\subsection{The DW correlators}

Here, we provide the derivation of the evolution of the various expectation values and correlators utilized above. We start with
\begin{align}
    \frac{d}{dt}\langle A_{jk}(t)\rangle&=i(\epsilon_j-\epsilon_k)\langle A_{jk}(t)\rangle-\sum_{j',k'}(\Gamma_{j'k'}^n+\gamma_{j'k'}){\rm Tr}\{ A_{jk}(A_{k'j'}A_{j'k'}\rho_\Phi+\rho_\Phi A_{k'j'}A_{j'k'})\}\nonumber\\
    &+2\sum_{j',k'}(\Gamma_{j'k'}^n+\gamma_{j'k'}){\rm Tr}\{A_{k'j'}A_{jk}A_{j'k'}\rho_\Phi \}=[i(\epsilon_j-\epsilon_k)-\underbrace{\sum_{p}(\Gamma_{pk}^n+\gamma_{pk}+\Gamma_{pj}^n+\gamma_{pj})]}_{\displaystyle \Gamma_{jk}^{\rm eff}}\langle A_{jk}(t)\rangle\,,
\end{align}
so that 
\begin{align}
    \langle A_{jk}(t)\rangle=e^{i(\epsilon_j-\epsilon_k)t}e^{-\Gamma_{jk}^{\rm eff}t}\langle A_{jk}(0)\rangle\,.
\end{align}
Hence, we can use the quantum regression theorem \cite{gardiner00} to extract the correlators:
\begin{align}
    \langle A_{jk}(t+\tau)A_{j'k'}(t)\rangle=e^{i(\epsilon_j-\epsilon_k)\tau}e^{-\Gamma_{jk}^{\rm eff}\tau}\langle A_{jk}(t)A_{j'k'}(t)\rangle\equiv e^{i(\epsilon_j-\epsilon_k)\tau}e^{-\Gamma_{jk}^{\rm eff}\tau}p_j\,,
\end{align}
where in the last line we assumed the long-time (or stationary) limit density matrix $\rho_\Phi=\sum_j p_j|j\rangle\langle j|$.

\end{widetext}


\begin{thebibliography}{31}
\bibitem [{rem()}]{remark}%
  \BibitemOpen
  \bibinfo {note} {The presence of a magnetic field and easy-axis anisotropy
  causes extra DW-magnon interactions of Hamiltonian origin, but they scale as
  $\propto1/L$ \cite{SM}, and hence are subleading as compared to the Berry
  phase contribution.}\BibitemShut {Stop}%
\end{thebibliography}

\end{document}